    \documentclass[12pt,draftcls,peerreview,letterpaper,onecolumn]{IEEEtran}

\usepackage{setspace}
\usepackage{graphics}
\usepackage{psfrag}
\usepackage{amsmath}
\usepackage{amsfonts}
\usepackage{amssymb}
\usepackage{mathrsfs}
\usepackage{graphicx}
\usepackage{subfigure}
\usepackage[top=0.65in, bottom=0.65in, left=0.86in, right=0.86in]{geometry}
\usepackage{fixltx2e}
\usepackage{setspace}
\begin{document}
\title{Analysis of Blocking Probability in a Relay-based Cellular OFDMA Network
}
% Information Networks Lab, Department of Electrical Engineering\\
\author{\authorblockN{Mahima Mehta, Ranjan Bala Jain and Abhay Karandikar\\ }
\authorblockA{Information Networks Lab, Department of Electrical Engineering\\Indian Institute of Technology Bombay, Mumbai 400 076, India
\\ Email:\{mahima, rbjain, karandi\}@ee.iitb.ac.in} \\}

\maketitle
%\doublespacing
\begin{abstract}
Relay deployment in Orthogonal Frequency Division Multiple Access (OFDMA) based cellular networks helps in 
coverage extension and/or capacity improvement. \textbf{To quantify capacity improvement, blocking probability of voice traffic is typically calculated using Erlang B formula. This calculation is }
% To analyze the capacity improvement with relay deployment, we 
% evaluate the Erlang capacity or blocking probability for such networks.  
% In general, the Erlang B formula is used to calculate the blocking probability of voice traffic, which is
based on the assumption that all users require same amount of resources to satisfy their rate requirement. However, in an OFDMA system, each 
user requires different number of subcarriers to meet its rate requirement. This resource requirement depends on 
the Signal to Interference Ratio (SIR) experienced by a user. 
Therefore, the Erlang B formula can not be employed to compute blocking 
probability in an
%Long-Term Evolution (LTE) networks which deploy 
OFDMA network. \textbf{In this paper, we determine an analytical expression to 
compute the blocking probability of relay based cellular OFDMA network. We determine an expression of the probability distribution of the user's resource requirement
%that discrete random variable which indicates the number of resources required by a user, 
based on its experienced SIR. Then, we classify the users into various classes depending upon their subcarrier 
requirement. }We consider the system to be a multi-dimensional system with different classes and
%Then, we model the system states by a multi-dimensional Markov chain and finally, 
evaluate the blocking probability of system using the
multi-dimensional Erlang loss formulas. \textbf{
This model is useful in the performance evaluation, design, planning of resources and call admission control of relay based cellular OFDMA networks like LTE.}	
% \cite{multi}.
\let\thefootnote\relax\footnotetext{Part of this paper (Section-III) is presented in an Int. Conf. on Information Processing (ICIP), 2011, Bangluru, India.} %(Published by Communications in Computer and Information Science Series, Springer Computer Science, Vol.157, Aug. 2011) and published in an Int. Journal of Computer Applications (IJCA), Special Issue on Wireless Communication and Mobile Networks, Jan. 2012, Foundation of Computer Science, New York, USA.}
\end{abstract}
%\begin{keywords}
% Blocking Probability, Inter-Cell Interference, Markov Chain Modeling, %Multi-Dimensional System.
%\end{keywords}
\section{Introduction}
The Third Generation Partnership Project Long-Term
Evolution (3GPP-LTE) proposes different schemes for mobile broadband access in order to meet the throughput and coverage requirements of
next generation cellular networks \cite{LTE}. Deployment of Relay Stations (RSs) to increase coverage area and/or improve capacity \cite{Throughput} is one of the proposed techniques in LTE. %The transmit power level of RS is lower than that of the Base Station (BS) and the radio resources are shared between the two. 
In this paper, we analyze the capacity improvement due to RS deployment and analytically determine the blocking 
probability to quantify this improvement. Blocking probability corresponds to the probability that a user is denied sevice due to non-availability of sufficient resources in the network.\\
\indent Users who experience poor signal strength from Base Station (BS), require more resources to meet 
their rate requirement and a large amount of resources are consumed in serving such 
users. This leads to an increase in the blocking probability. With RS deployed in the network, the Signal to Noise Ratio (SNR) experienced 
by these users may improve due to closer proximity of RS and as a result, they may meet their rate requirement with fewer 
resources. This reduces the blocking probability and improves the system capacity. However, as the radio resources are shared between BS and RS, deployment of RSs introduce additional sources of interference. Therefore, it is significant to study the impact of interference on the blocking probability.\\
\indent Blocking probability has been used as a performance metric in \cite{QOS_guarantee}, in which the transmission 
scheme selection policy (single hop or multi-hop) has been proposed to provide guaranteed target Bit Error Rate 
(BER) and data rate to a mobile user. Another metric to quantify the performance improvement in a cellular network 
is Erlang capacity, which is the traffic load in Erlangs supported by the cell while ensuring that blocking 
probability remains less than a certain value. There is sufficient work
%several papers 
on Erlang capacity and blocking probability in cellular networks \cite{GSM_blocking}, \cite{CDMA_blocking} and some literature is available on determining the Erlang capacity of cellular Orthogonal Frequency Division Multiple Access (OFDMA) networks.
In \cite{Reviewer_1}, the performance of subcarrier allocation in OFDM system has been investigated considering multi-class users. However, in this work, the subcarriers are not released simultaneously, as would happen in practice but are released one by one. The Erlang Loss Model for blocking probability analysis has been suggested in \cite{Reviewer_2} and is proved to be numerically efficient and insensitive to the distribution of call duration. More recently in \cite{Reviewer_3}, the OFDM system for blocking probability computation considers power and subcarrier allocation for users. %However, the justification of the number of subcarriers required to achieve certain data rate by a user is not given.
Despite the availability of sufficient literature on determining the Erlang capacity and blocking probability in cellular networks including OFDMA systems, limited literature is available on determining the Erlang capacity of relay 
based cellular OFDMA networks \cite{uplink2}, \cite{amc}, \cite{Gauri}. In \cite{uplink2}, the uplink 
Erlang capacity of relay-based OFDMA network has been derived considering adaptive modulation and coding supporting both 
voice and data traffic. In \cite{amc}, the uplink capacity and spectral efficiency of relay-based cellular networks have been analyzed. The bandwidth distribution between BS and RSs has been determined to ensure that the blocking probability is less than a specific threshold. The impact of number of RSs and their positions on Erlang capacity is investigated by considering Adaptive Modulation and Coding (AMC) and Multiple-Input Multiple-Output (MIMO) transmissions. \\
\indent In \cite{uplink2} and \cite{amc}, it is assumed that all users require equal number of resources. 
However, the impact of user location, shadowing and interference from neighboring cells on the resource requirement has not been considered. If distinct users of same data rate requirement are present at different locations, they may experience different Signal to Interference Ratio (SIR) and hence require different resources in terms of number of subcarriers to satisfy their data rate requirement. In the queuing literature, the problem of incoming users requiring different number of resources has been addressed in some works. %\cite{Glimpson} and \cite{Kaufman}.
In \cite{Glimpson}, wide band and narrow band traffic is considered, where no queuing is allowed for narrow band traffic and a finite length queue is provided for wide band traffic. The blocking probability for each traffic class is determined using numerical methods. Similarly, in \cite{Kaufman} and \cite{multi}, the problem of multiple server requirement is analyzed and multidimensional Erlang loss formulas have been derived. \\
%{ We have addressed this issue for relay-based OFDMA systems in this paper. We compute the SIR experienced by each user and accordingly determine the number of subcarriers required.}\\
%where it has been 
%shown that the steady state distribution of number of users of each traffic has a product form.\\
\indent To the best of our knowledge, no literature is available for the computation of blocking probability in relay based cellular OFDMA systems, where different users of same rate requirement need different subcarriers. In \cite{Gauri}, different subcarrier requirement of users has been considered in the blocking probability computations. However, SIR experienced by a user and distribution of subcarrier requirement were determined using simulations. In \cite{ICIP_Ranjan} (by one of the authors), Cumulative Distribution Function (CDF) of interference is computed analytically. However, blocking probability is not determined.\\
\indent In this paper, we propose an analytical model to evaluate the performance of a relay based cellular OFDMA network (such as an LTE network) in terms of blocking probability. The distinct feature of our paper is that we consider the impact of user location, shadowing and interference from neighboring cells in our analysis for blocking probability. Specifically, we determine the SIR experienced by a user and probability distribution of the number of subcarriers required. Then, we classify incoming users into different classes based on their subcarrier requirement. We consider the network to be a multi-dimensional system with different classes and model the system states by multi-dimensional Markov chain. In such a system model, the 
computational complexity is more due to the large state space involving the states of both BS and RS. To reduce this 
complexity, we propose an approximation where the state space of BS and RS are decoupled. With this simplification, we evaluate the blocking probability of each class in a relay based OFDMA system. This approximation is justified by comparing the analytical results with simulation results where we do not make such assumption.\\  
%a model where the states of subcarriers used at the BS and RS are defined separately. We model the system states by multi-dimensional Markov chain and evaluate the blocking probability of each class at BS and RS separately. Then, we evaluate an overall blocking probability of the system.}\\
\indent The rest of the paper is organized as follows. Section \ref{sysmodel} introduces the system model for the downlink of relay based cellular OFDMA network. % and outlines the problem statement.
In Section \ref{ICI}, a model to characterize Inter-Cell Interference (ICI) on a Mobile Station (MS) is presented 
and the CDFs of ICI on BS-MS, BS-RS and RS-MS transmission links are derived. In 
Section \ref{class}, an analytical model is proposed to determine the subcarrier requirement and its probability distribution based on ICI experienced. 
In Section \ref{blocking}, the incoming users are classified into various classes based on their subcarrier requirement. It is also shown that complexity is introduced due to the large size of state space when 
both BS and RS are considered. Then, an analytical model is developed by considering the state space of BS and RS
separately. This model is used to compute the blocking probability for each 
class of user at BS and RS. Finally, the blocking probability of a relay-based OFDMA network is computed using 
multi-dimensional Erlang loss formulas \cite{multi}.
In Section \ref{numericalresults}, the simulation methodology is explained and both analytical and simulation results are discussed. Here, the system performance (in terms of blocking probability) of a non-relay system with that of the relay-based cellular OFDMA system is compared. Finally, Section \ref{conclusion} concludes the paper with an insight into the future extensions of the present work.
\section{System Model}\label{sysmodel}
\begin{figure*}[h!]
  \begin{center}
\scalebox{0.5}{\input{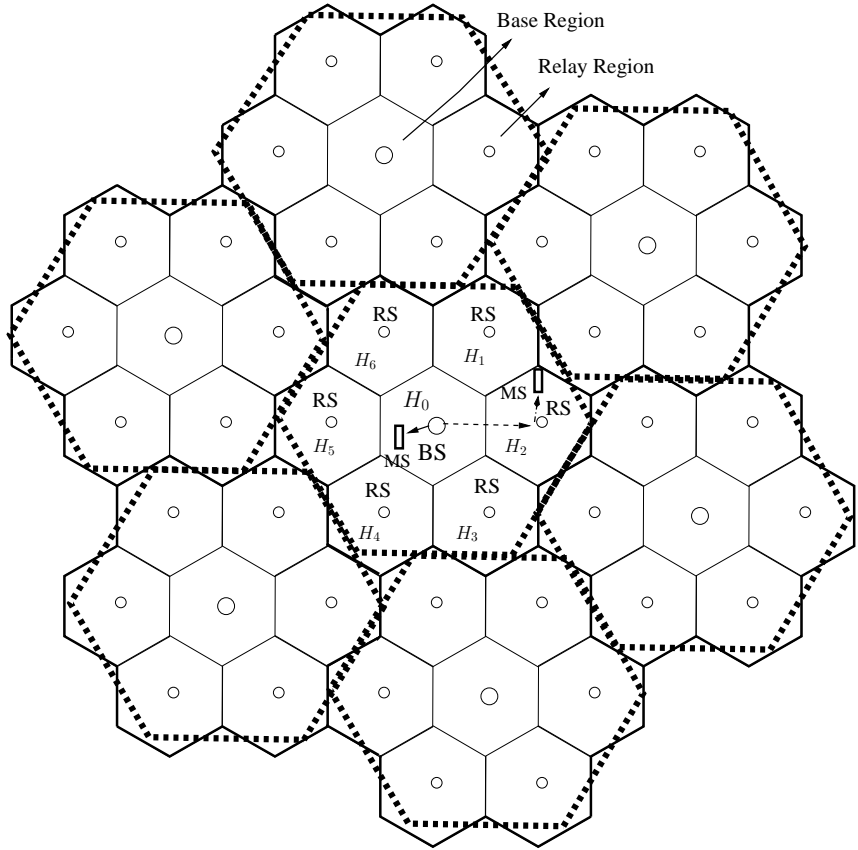}}
\center \caption{Architecture of Relay based Cellular OFDMA System}
  \label{cell_structure}
  \end{center}
  \end{figure*}

We consider the downlink transmission scenario in a relay-based cellular OFDMA network as shown in 
Fig.\ref{cell_structure}. We define the reference cell as a combination of seven sub-cells. The central 
sub-cell $H_{0}$ consists of a BS centered at $(0,0)$, while each surrounding sub-cell (i.e. $H_{1}$, $\ldots$ , 
$H_{6}$) consists of one RS at the centre. For convenience, we approximate the coverage of BS and RS by hexagons as shown in Fig.\ref{cell_structure}.
We define the central sub-cell $(H_{0})$ as base region and the six 
surrounding sub-cells $(H_{1}$, $\ldots$ , $H_{6})$ as relay region in every cell. 
We define a MS (user) present in the reference cell as target MS.
We assume that BS and RS have 
Line of Sight (LoS) connection. % and BS-MS and RS-MS links have non LoS. 
All RSs are assumed to be amplify-and-forward type relays. \textbf{However, we consider the factor of amplified noise to be small and therefore, neglect that in our calculations further.} %, which decode the data received from BS and then forward it to the target MS.
\\
\indent We consider universal frequency reuse, i.e. all cells use the same spectrum, which is shared between BS and six RSs. %However, universal frequency reuse introduces more severe ICI. 
We consider the interference from the first tier of neighboring cells only.
We also consider the effect of path loss and lognormal shadowing on the transmitted signal. Let the BS transmit at power 
$P$ to a MS located at distance $d$, then the received power at MS will be  $Pd^{-\beta}10^{\xi/10}$, where $\beta$
 is path loss exponent and $\xi$ represents lognormal shadowing on BS-MS link. $\xi$ is a Gaussian random variable with mean $0$ dB and standard deviation $\sigma$. Since thermal noise is negligible in an interference-limited reuse-one network, we ignore it in our computations. Note that we do not consider fast fading and frequency-selective fading as our objective is to evaluate the blocking probability from a long term capacity planning perspective. For the same reason, we do not consider any power control mechanism and assume that BS transmits at fixed power. \\
\indent In practice, the association of MS with BS or RS is determined based on SIR. If SIR experienced by MS from BS is above threshold, then it will be associated with BS otherwise with RS. In the present paper however, we consider a model where users present in BS region are associated with BS directly and those present in relay region are associated with the corresponding RS. It is assumed that the users are uniformly distributed in the respective regions of the cell. As explained later in Section \ref{blocking}, out of the total call arrivals to the cell, a fraction is assumed to occur in BS region, while the remaining are assumed to have occured in relay region. 
% path loss to dominate in the base region and therefore, assume that if MSs are located in base region, they will be associated with the BS. However, we consider both path loss and shadowing in relay region and therefore, if MSs are located in any of the relay regions, they will be associated with the corresponding RS or BS depending on the received SIR at that MS. 
% If SIR experienced by MS from BS is above threshold, then it will be associated with BS directly otherwise with RS. 
% If SIR experienced by MS from both BS and RS is above threshold , then BS is preferred over RS for association with the MS.
% =======================================
% =======================================
\\ \indent
%We consider a system where the resources are partitioned in the time-frequency band as shown in Fig.\ref{Txn_sys}. The x-axis and y-axis represent time slots and frequency bands respectively. When the transmission takes place on any of the links (BS-MS, BS-RS and RS-MS), a set of  sub-carriers are assigned to a user for certain time slots. Here, we assume that  for a direct call BS-MS transmission takes place in complete time slot from BS to MS. However, each time slot is partitioned into two for a hopped call as shown in the figure. BS-RS transmission takes place in first half of the slot and RS-MS in the second half. 
% =============
%\let\thefootnote\relax\footnotetext{In a practical cellular system like LTE, specific subframes known as the Multicast/Broadcast Single Frequency Network (MBSFN) subframes \cite{LTE_RS_Txn} are utilized for transmission on BS-RS link and other subframes are used for transmission on RS-MS link. Though we have not specifically considered this practical scenario, our system model captures such transmission scenario.}
We assume that there are $K$ number of subcarriers available in the reference cell, which are shared between BS and six RSs. Each RS and BS are allocated $K_{RS}$ and $K_{BS}$ subcarriers respectively, such that $K = K_{BS} + 6 K_{RS}$. %This subcarrier sharing is represented in Fig.\ref{Txn_sys} on the vertical axis above ($K_{BS}$) and below ($6 K_{RS}$) the horizonal axis. 
If insufficient number of subcarriers are allocated to RS, then RS will not be able to relay the signals received from BS to MS. On the other hand, if the subcarriers allocated to RSs are more than the required, then there may be an increase in call blocking at the BS. Thus, the value of $K_{RS}$ may influence the overall system performance and therefore needs to be carefully designed. In this paper, we do not consider an optimal method for sharing the subcarriers amongst BS and six RSs. \\
%\indent \textbf{We consider the system to be fully loaded (i.e. all $K$ subcarriers are in use in all neighboring cells). A user present at a distance $d$ from the serving BS will experience that the total interference power from all neighboring cells on each subcarrier is equal. Since SIR on each subcarrier is same and we do not consider Rayleigh fading (for blocking probability analysis), there is no need for power control mechanism in the system. Therefore, we consider equal power allocation on all subcarriers.}
% We also assume that RS uses different frequency channels to communicate to BS and MSs. 
%does its transmission and reception in different frequency band. 
\indent In the reference cell, all users have been 
allocated orthogonal subcarriers and therefore no intra-cell interference exists. However, in a network with universal
frequency reuse, users will experience interference from RSs and BSs of neighboring cells. We consider the system to be fully loaded (i.e. all $K$ subcarriers are in use in all neighboring cells
%) and the interference only from the 
of the first tier of cells). We analyze this interference on BS-MS, BS-RS and RS-MS link and compute their Cumulative Distribution Functions (CDFs).
Using these CDFs, we determine the probability distribution of subcarrier requirement on these three links. 
% Then, incoming users are classified into various classes depending on their subcarrier requirement, to achieve the desired rate.
 We consider
the rate requirement to be same for all users. The blocking probability on each link is calculated 
and then overall blocking probability for relay based OFDMA network is determined.
\section{Inter-Cell Interference modeling}
\label{ICI}
In this section, we consider a target MS (user) in the reference cell.
%, which can be located in either the base region or relay region. 
We analyze the SIR experienced by the target MS on BS-MS link if it is associated with BS and on BS-RS and RS-MS links, if it is associated with RS. Then, we compute the CDF of SIR on these links following \cite{ICIP_Ranjan}. For this, we divide the incoming users into two groups,\\
 Group 1: Users present in base region are associated with the BS directly on the BS-MS link. These users are called direct users.\\
 Group 2: Users present in the relay regions are associated with the BS via corresponding RS on BS-RS and RS-MS links. These users are called hopped users.
\\Note that we use the terms users and calls interchangeably in this paper. % and the term MS refers to `target' user.
\subsection{ SIR on BS-MS, BS-RS and RS-MS transmission links }
Let $\gamma_{BS-MS}$, $\gamma_{BS-RS}$ and $\gamma_{RS-MS}$ denote SIR on a subcarrier used on BS-MS, BS-RS and RS-MS links respectively. Then we have,
\begin{equation}
\gamma_{BS-MS}  = \frac{P_{BM} d_{BS-MS}^{-\beta} 10^{\frac{\xi_{BS-MS}}{10}}}{\sum_{i=1}^{N}P_{BM}d_{i BS-MS}^{-\beta} 10^{\frac{\xi_{i BS-MS}}{10}} },
\label{snr_bs}
\end{equation}
\begin{equation}
\gamma_{BS-RS}  =  \frac{P_{BR} d_{BS-RS}^{-\beta} 10^{\frac{\xi_{BS-RS}}{10}}}{\sum_{i=1}^{N}P_{BR} d_{i BS-RS}^{-\beta} 10^{\frac{\xi_{i BS-RS}}{10}} }
\label{snr_rs}
\end{equation}
and
\begin{equation}
\gamma_{RS-MS}  =  \frac{P_{RM} d_{RS-MS}^{-\beta} 10^{\frac{\xi_{RS-MS}}{10}}}{\sum_{i=1}^{N}P_{RM} d_{i RS-MS}^{-\beta} 10^{\frac{\xi_{i RS-MS}}{10}} },
\label{snr_ms}
\end{equation}
where, \begin{itemize}
\item $P_{BM}$, $P_{BR}$ and $P_{RM}$ denote the  power transmitted by BS to target MS, BS to RS and RS to target MS respectively. 
\item $d_{BS-MS}$, $d_{BS-RS}$ and $d_{RS-MS}$ denote the distance between BS and target MS present in base 
region, BS and RS (with which the target MS is associated) and RS and target MS present in any of the relay regions respectively.
\item $d_{i BS-MS}$, $d_{i BS-RS}$ and $d_{i RS-MS}$ denote the distance between $i^{th}$ neighboring BS and target MS (present in base region), $i^{th}$ neighboring BS and RS (with which the target MS is associated) and $i^{th}$ neighboring RS and target MS (present in any of the relay regions) respectively.
% denote the distance as mentioned above from $i^{th}$ neighboring BS and RS to respective target MS and RS in reference cell.
% between $i^{th}$ neighboring BS (which is using same subcarrier as used by the reference BS) and MS present in base region, $i^{th}$ neighboring BS and RS (who is serving the MS) and RS (which is using same subcarrier as used by the RS of reference BS) of $i^{th}$ BS and MS present in one of the relay regions respectively.
\item $N$ is the number of interferers in the first tier of cells.
\item $\xi_{BS-MS}$, $\xi_{BS-RS}$ and $\xi_{RS-MS}$ represent lognormal shadowing on BS-MS, BS-RS and RS-MS links. Each of them is a Gaussian random variable with mean $0$ dB and standard deviations $\sigma_{BS-MS}$, $\sigma_{BS-RS}$ and $\sigma_{RS-MS}$ dB respectively.
\item $\xi_{i BS-MS}$, $\xi_{i BS-RS}$ and $\xi_{i RS-MS}$ represent lognormal shadowing on $i^{th}$ neighboring BS and target MS link, $i^{th}$ neighboring BS and RS link, and $i^{th}$ neighboring RS and target MS link. Each of them is a Gaussian random variable with mean $0$ dB and standard deviations $\sigma_{i BS-MS}$, $\sigma_{i BS-RS}$ and $\sigma_{i RS-MS}$ dB respectively.
\end{itemize}
\subsection{CDF of SIR }
\label{cdf_BS_MS}
In this section, we determine mean and variance of interference to signal ratio $I_{BS-MS}$ in two steps.
In \cite{Filho}, \cite{Yeh} and \cite{Kostic}, it has been argued that the total interference power received from 
various interferers (in a universal frequency reuse system) can be modeled by lognormal distribution with some mean and variance. We make the same assumption here. Accordingly, we proceed to calculate the mean and variance of $I_{BS-MS}$.\\ We rewrite Eq. \ref{snr_bs}  as,
\begin{eqnarray*}
\gamma_{BS-MS} & = & \frac{1}{\sum_{i=1}^{N}(\frac{d_{i BS-MS}}{d_{BS-MS}})^{-\beta} 10^{\frac{\xi_{i BS-MS}-\xi_{BS-MS}}{10}}}=  \frac{1}{I_{BS-MS}}.
\end{eqnarray*}
% \begin{enumerate}
% \item 
\textbf{Step-1:} Let $(0,0)$, $(x, y)$ and $(x_{i}, y_{i})$ be the coordinates of BS in reference cell, target MS in 
reference cell and the $i^{th}$ interfering BS present in the first tier respectively. \\ $I_{BS-MS}$ is grouped into two components, $B_{i}$s and $C_{i}$s as,
\begin{equation}
I_{BS-MS} = \sum_{i=1}^{N}B_{i} C_{i},
\end{equation}
where, $
%\begin{eqnarray*}
B_{i}= \left( \frac{d_{i BS-MS}}{d_{BS-MS}}\right) ^{-\beta}=\left[ \frac{(x-x_{i})^{2}+(y-y_{i})^{2}}{x^{2}+y^{2}}\right] ^{\frac{-\beta}{2}} $
%\end{eqnarray*} 
and
%\begin{eqnarray*}
$C_{i}=10^{\frac{\xi_{i BS-MS}-\xi_{BS-MS}}{10}}$
% \frac{\xi_{i BS-MS}}{\xi_{BS-MS}}$.\\
%\end{eqnarray*}
\\ $B_{i}$ is the ratio of distances and is a function of the position $(x, y)$ of user in the reference cell. The position of target MS in the reference cell is random  but the interfering BSs have fixed positions. Therefore, $d_{i}$s are correlated and as a result, $B_{i}$s are correlated RVs.
%
%The position $(x, y)$ of target MS in the reference cell is random but the interfering BSs have fixed positions. Therefore, $d_{i BS-MS}$s are correlated and as a result, $B_{i}$s are correlated random variables. 
\\ \indent $C_{i}$ is a ratio of two lognormal RVs, shadowing from $i^{th}$ interfering BS to the target MS and 
shadowing from the serving BS to the target MS. As suggested in \cite{son}, it can be approximated by a lognormal RV with mean $0$ and variance 
$(\sigma_{i BS-MS}^{2} + \sigma_{ BS-MS}^{2})$. Thus, all $C_{i}$s are lognormal RVs but correlated. Note that lognormal shadowing $\xi$ is independent of position of user. Hence, it is reasonable to also assume for 
% Shadowing represents the obstructions in environment and it is independent of position of user. We assume shadowing to be independent of user distance, hence $B_{i}$ and $C_{i}$ are independent. We also assume that 
$B_{i}$ and $C_{j}$ to be independent for any pair $(i,j)$. \\
We assume that,\\ $E\left[C_{i}\right]= E\left[C_{j}\right]$ and $E\left[C_{i}C_{j}\right]$ = constant,  $\forall$ $i \neq j $.
\\ The first and second moments of $I_{BS-MS}$ are determined as,
\begin{equation}
E[I_{BS-MS}] = E\left[ C_{i}\right] E\left[ \sum_{i=1}^{N}B_{i}\right],
\label{mean}
\end{equation}
\begin{equation}
\begin{split}
E\left[ I_{BS-MS}^{2}\right]  = E\left[ C_{i}^{2}\right] E\left[ \sum_{i=1}^{N}B_{i}^{2}\right] + E\left[ C_{i}C_{j}\right] \left[ E\left( \sum_{i=1}^{N}B_{i}\right) ^{2}-E\left( \sum_{i=1}^{N}B_{i}^{2}\right) \right].
\end{split}
\label{var}
\end{equation}
In Eq. \ref{mean} and \ref{var}, computations of $E[C_{i}]$, $E[C_{i}^{2}]$ and $E\left[ \sum_{i=1}^{N}B_{i}\right]$ are straightforward. $E\left[ \sum_{i=1}^{N}B_{i}\right] ^{2}$ and $E\left[ \sum_{i=1}^{N}B_{i}^{2}\right] $ are solved as follows,
\\ Since the distances $d_{i BS-MS}$s between the target MS and the $i^{th}$ interfering fixed BS are correlated, $E\left[\sum_{i=1}^{N}B_{i}\right]^{2}$ can not be separated into a sum of terms. It is computed by averaging over the area as follows,
%Let, 
% \fontsize{10}
\begin{align}\label{eq7}
 E\left[\sum_{i=1}^{N}B_{i}\right]^{2} &=
% \end{equation*}
% \begin{eqnarray}%A_{s} = 
\dfrac{2}{3\sqrt{3}} \iint\limits_{x,y\in H_{0}}\left[\sum_{i=1}^{N} \sqrt{\dfrac{(x-x_{i})^{2}+(y-y_{i})^{2}}{{x^{2}+y^{2}}}}^{ -\beta}\right] ^{2}  dx dy.
\end{align}
% \begin{align}
% A_{s} &= \nonumber\\
% &\frac{2}{3\sqrt{3}} \iint\limits_{x,y\in H_{0}}\left[\sum_{i=1}^{N} \sqrt{\frac{(x-x_{i})^{2}+(y-y_{i})^{2}}{{x^{2}+y^{2}}}}^{ -\beta}\right] ^{2}dx dy.
% \end{align}
Now, to compute $E\left[ \sum_{i=1}^{N}B_{i}^{2}\right]$, %$K = \frac{(x-x_{i})^{2}+(y-y_{i})^{2}}{{x^{2}+y^{2}}}$ and%
expectation is taken over all possible positions  $(x, y)$ the target MS can take
% as it is uniformly distributed 
in the base region.
%$H_{0}$ centered at $(0, 0)$. The area of each cell is $3\sqrt{3}/2$. %
These integrals are evaluated separately for each interfering BS and then summed for all BSs to
get $E\left[ \sum_{i=1}^{N}B_{i}^{2}\right] $, as shown below,
\begin{eqnarray}
%M &=& 
\label{eq8}
E\left[ \sum_{i=1}^{N}B_{i}^{2}\right] = \sum_{i=1}^{N}E \left[B_{i}^{2}\right] = 
% & \\ \nonumber
% &=& \sum_{i=1}^{N}\frac{2}{3\sqrt{3}} \iint_{x,y\in H_{0}}\left(\sqrt{\frac{(x-x_{i})^{2}+(y-y_{i})^{2}}{{x^{2}+y^{2}}}}\right)^{-2\beta} dx dy
% \end{eqnarray}
\sum_{i=1}^{N}\frac{2}{3\sqrt{3}} \iint \limits_{x,y\in H_{0}}\left[\frac{(x-x_{i})^{2}+(y-y_{i})^{2}}{x^{2}+y^{2}}\right]^{-\beta} dx dy.
\end{eqnarray}
Eq. \ref{eq7} and \ref{eq8} can be solved numerically for hexagonal geometry.\\
 \textbf{Step-2:} We have obtained first and second moments of $I_{BS-MS}$ in Step-1 (Eq. \ref{mean} and \ref{var}). Its distribution can be approximated by lognormal distribution with parameters $(\mu_{I_{BS-MS}},\sigma_{I_{BS-MS}}^{2})$. \\ 
In general the $k^{th}$ moment can be written as,
\begin{equation}
E\left[ I^{k}_{BS-MS}\right]  = e^{k \mu_{I_{BS-MS}}+\frac{k^{2}}{2}\sigma^{2}_{I_{BS-MS}}}.
\end{equation}
%where, $k$ is order of moments. 
Using $k$ = $1$ and $2$ and on inverting, we obtain,
\begin{equation}
\mu_{I_{BS-MS}}= 2 ln E[I_{BS-MS}] - \frac{1}{2} ln E[I_{BS-MS}^{2}]\label{mean2}
\end{equation}
and
\begin{equation}
\sigma_{I_{BS-MS}}^{2}= -2 ln E \left[ I_{BS-MS}\right] + ln E \left[ I_{BS-MS}^{2}\right].\label{var2}
\end{equation}
Using Eq. \ref{mean2} and \ref{var2}, we determine the distribution as,
% $\mu_{I_{BS-MS}}$ and $\sigma_{I_{BS-MS}}^{2}$, we determine the distribution as,
\begin{equation}
\label{F_bs_ms}
F_{\mathbf{I_{\mathbf{BS-MS}}}}(x)= \Phi\left[\frac{lnx- \mu_{I_{BS-MS}}}{\sigma_{I_{BS-MS}}}\right], x > 0.
\end{equation}
Here $\Phi(x)$ is the standard normal CDF.\\
% \end{enumerate}
Similar calculations are performed to obtain the CDF of $I_{BS-RS}$ and $I_{RS-MS}$ on BS-RS and RS-MS links as, 
\begin{equation}
F_{\mathbf{I_{\mathbf{BS-RS}}}}(x)= \Phi\left[\frac{lnx- \mu_{I_{BS-RS}}}{\sigma_{I_{BS-RS}}}\right], x > 0
\end{equation}
and
\begin{equation}
F_{\mathbf{I_{\mathbf{RS-MS}}}}(x)= \Phi\left[\frac{lnx- \mu_{I_{RS-MS}}}{\sigma_{I_{RS-MS}}}\right], x > 0.
\end{equation}
Thus, we have determined the distribution of interference to signal ratio on a subcarrier on the three
transmission links, i.e., BS-MS, BS-RS and RS-MS links.
\section{Analytical Model to determine resource requirement based on CDF of SIR }\label{class}
In cellular OFDMA system, an incoming user is allocated a certain number of sub-carriers to satisfy its rate 
requirement. In our formulation, we consider that all incoming users have the same rate requirement $R$. 
Due to different SIR experienced by the users, they will require different number of subcarriers. \\
% % In this section, we propose a model to calculate the number of subcarriers required by each user, determine their probability distribution and then classify them into different classes based on their subcarrier reqiurement.\\
% \subsection{Determining Number of Subcarriers and their Probability Distribution}
% \label{determination1}
\indent The objective of BS is to satisfy the rate requirement of each user, by 
allocating it the requested number of subcarriers which depends upon its experienced SIR. There are $K_{BS}$ orthogonal subcarriers available at the BS, each of bandwidth $W$ Hz. \\
\indent Let, $\gamma^{m}_{BS-MS}$ be the SIR experienced by a user while using $m^{th}$ subcarrier on $BS-MS$ link. Then, the
rate $R$ achieved using $\mathcal{M}$ number of subcarriers on BS-MS link is given by,
\begin{equation}
R= W \sum_{m=1}^{\mathcal{M}}log_{2} \left(1+\gamma^{m}_{BS-MS}\right).
\end{equation}
%As discussed in Section \ref{sysmodel}, 
Since no frequency dependent fast fading is considered (Section \ref{sysmodel}), SIR on each subcarrier is same, i.e. $\gamma^{1}_{BS-MS}=\gamma^{2}_{BS-MS} \ldots \gamma^{M}_{BS-MS} = \gamma_{BS-MS} $, the number of subcarriers ($\mathcal{M}$) required by any user can be expressed as,
\begin{equation}
\label{number}
\mathcal{M}= \frac{R \cdot log_{10}(2)}{W \cdot log_{10}\left(1+\gamma_{BS-MS} \right)}.
\end{equation}
\indent Now, we distinguish users based on their subcarrier requirement as follows.
We divide the entire interference to signal ratio $I_{BS-MS}$ (determined in section \ref{cdf_BS_MS}) range into 
$l+1$ non-overlapping consecutive intervals with boundaries denoted by 
$\left\lbrace I^{r}_{BS-MS}\right\rbrace_{r=1}^{l+1}$. For each new user on BS-MS link, when the received interference to signal ratio $I_{BS-MS}$ falls in the range  $I_{BS-MS}\in\left[ I^{r}_{BS-MS}, I^{r+1}_{BS-MS}\right]$, then user is considered to be in 
class $r$. As $r$ lies in the range $(1,\ldots,l+1)$ %and according to the definition of class, 
the highest possible class of a user will be $l$ when 
$I_{BS-MS}$ falls in the range  $I_{BS-MS}\in\left[ I^{l}_{BS-MS}, I^{l+1}_{BS-MS}\right]$.\\
\indent Let $M^{r}$ denote the number of subcarriers required by the class $r$ user on that link. In the present case, we have $M^r = r$. 
%Thus, we consider that the number of subcarriers required $M^r$ is equal to the class $r$ itself. \\
% We associate each $r^{th}$ interval to the number of subcarriers $M^{r}$ such that $M^r = r$.\\
 Let $A = \frac{R}{W}log_{10}\left(2\right)$. Then, Eq. \ref{number} can be re-written as,
\begin{equation}
\label{boundary1}
I^{r}_{BS-MS} = (10^{\frac{A}{M^{r}}}-1)^{-1}.
\end{equation}
Note that Eq. \ref{boundary1} is used to determine the interference boundaries by assigning the number of subcarriers $M^r$ = $1, 2, \cdots,l$ to each $r^{th}$ interval $(r =  1, . . . , l)$. Thus, for each interval $(r = 1, \cdots , l)$ and assigned number of subcarriers  $(M^r = 1, 2, . . . ,l)$, the interference boundaries 
are determined.
%Thus, for each interval $n = 0,1, \ldots, l $  and corresponding number of subcarriers $M^{n}=1,2, \ldots, L $, the boundaries ${I^{n}_{BS-MS}}$ of BS-MS link are determined.\\
\indent Let, $ \mathbb{P}_{BS-MS}\left( {M^{r}}\right) $ denote the probability that an incoming user belongs to 
class $r$ and requires $M^{r}$ number of subcarriers on BS-MS link to meet its rate requirement. It is determined as,
%\begin{eqnarray}
%\mathbb{P}_{BS-MS}\left( {M^{n}}\right) & = & \text{Prob}\left[I^{n}_{BS-MS}< I_{BS-MS}< I^{n+1}_{BS-MS}\right]  \\ \nonumber
%& = &  F_{\mathbf{I_{BS-MS}}}\left( I^{n+1}_{BS-MS}\right)- F_{\mathbf{I_{BS-MS}}}\left( I^{n}_{BS-MS}\right). 
%\end{eqnarray}
\begin{align} 
\mathbb{P}_{BS-MS} \left( {M^{r}}\right) & = \label{Eq_18}
 \mathbb{P}\left[I^{r}_{BS-MS}< I_{BS-MS}< I^{r+1}_{BS-MS}\right]\\ \nonumber
& = F_{\mathbf{I_{BS-MS}}}\left( I^{r+1}_{BS-MS}\right)- F_{\mathbf{I_{BS-MS}}}\left( I^{r}_{BS-MS}\right). 
\end{align}
where $F_{\mathbf{I_{BS-MS}}}\left( I^{r}_{BS-MS}\right) $ is the CDF of interference to signal ratio 
(Eq. \ref{F_bs_ms}) on BS-MS link.
Similar calculations are performed to determine the \textbf{`probability of subcarrier requirement'}\footnote{\textbf{Note that the `probability of subcarrier requirement' represents the probability that a discrete random variable, say $x$ equals the number of subcarriers required by a user. For the ease of discussion, we use this terminology throughout this paper.}} on BS-RS and RS-MS links 
and they are denoted by $\mathbb{P}_{BS-RS}\left( {M^{r}}\right)$ and $\mathbb{P}_{RS-MS}\left( {M^{r}}\right)$ respectively. 
%\textbf{Note that we consider decode-and-forward type of relays, in which the achievable rate is the minimum of the rates achieved on BS-RS and RS-MS links. However, we allocate subcarriers to the users based on rate required $R_{req}$. Therefore, in our model the achievable rate with decode-and-forward type of relays will always be $R_{req}$.}
% \subsection{Classification of Users }
% \label{determination2}
% % \\ \indent Thus, we have determined the probability distribution of subcarrier requirement on all the 
% % three transmission links to meet the rate requirement $R$ of a user. We also classified the incoming 
% % users into different classes based on their subcarrier requirement.\\
% % \indent Now, we illustrate the notations that will be used in this paper to represent the number of subcarriers required by the users and their probability distribution. 
\section{Analysis of Blocking Probability}\label{blocking}
\begin{figure*}[h!]
\begin{center}
\scalebox{0.54}{\input{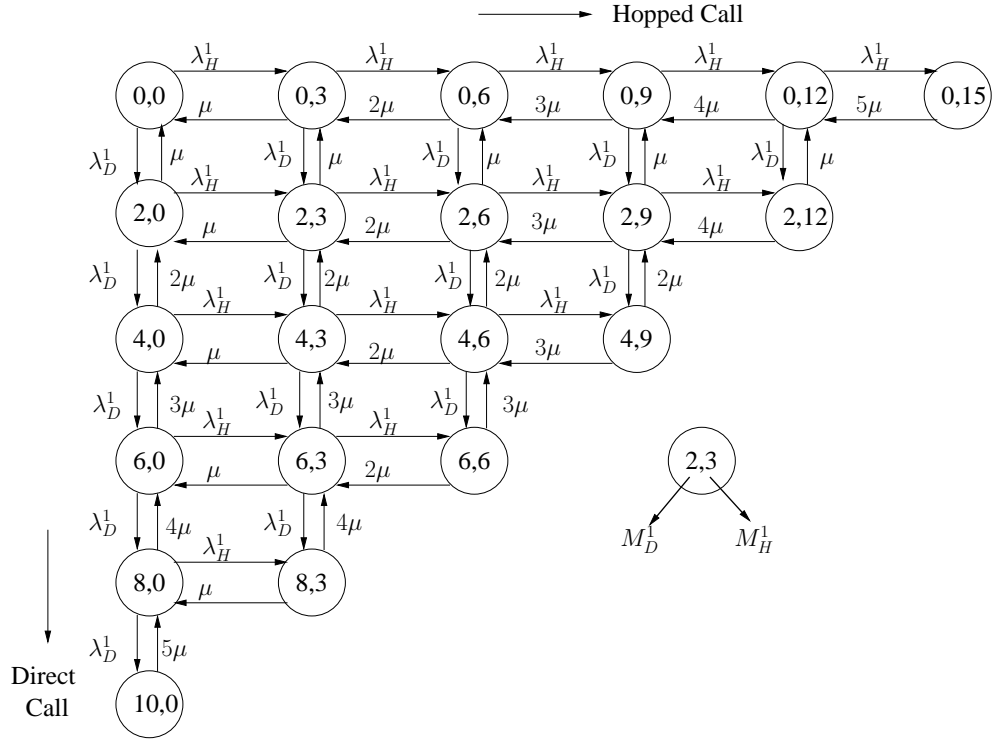}}
 \caption{Two-dimensional state transition diagram for a system with $K_{BS}=10$, $K_{RS}=6$, $M^{1}_{D}=2$ and $M^{1}_{H}=3$. Each call has only one class.}
\label{fig:markov1}
\end{center}
\end{figure*}

\begin{figure*}[h!]
\begin{center}
\scalebox{0.5}{\input{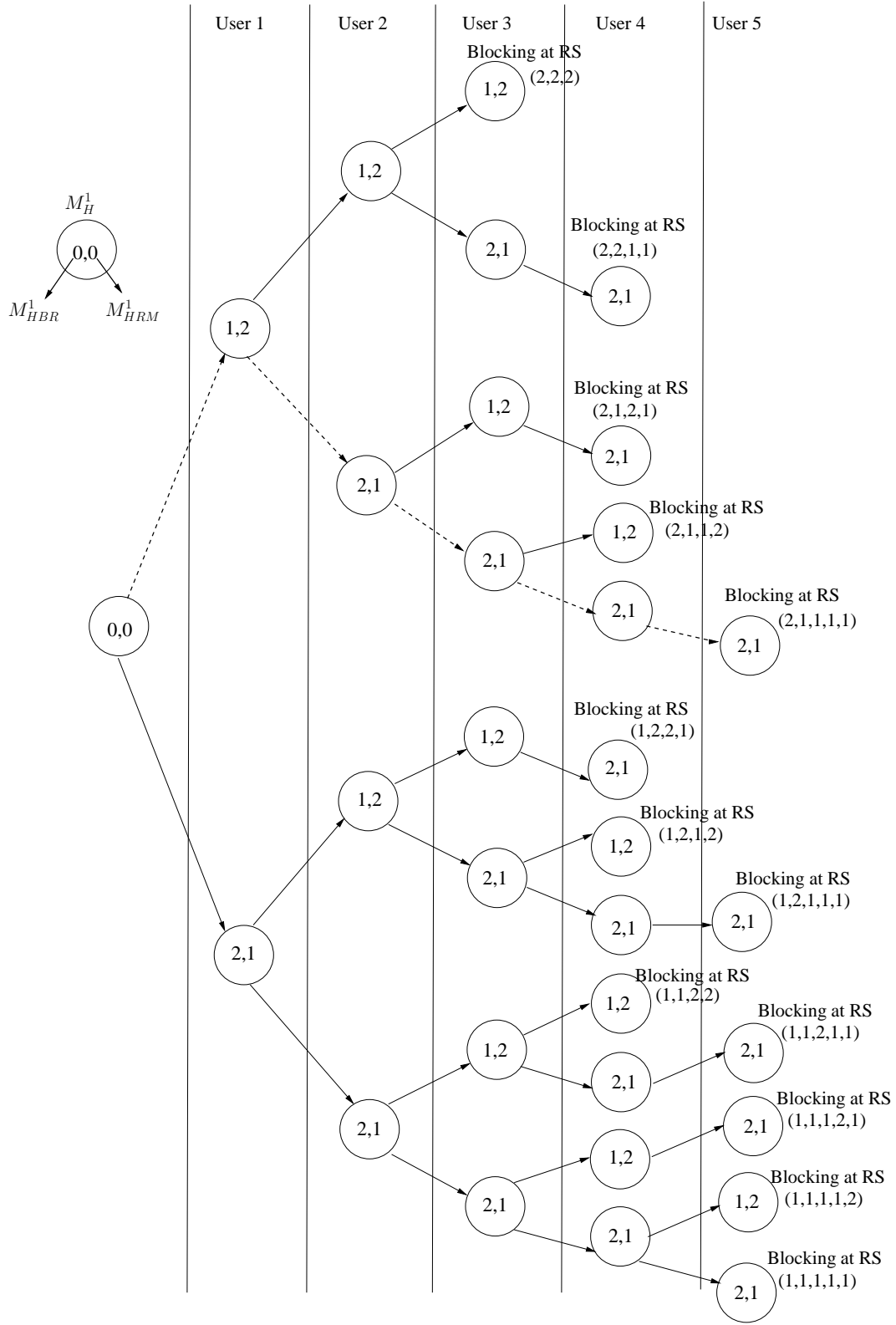}}
% \scalebox{0.9}{\includegraphics{markov2_new.pstex_t}}
\caption {Illustration of various possible combinations of subcarrier requirement on BS-RS and RS-MS links for a hopped call for a system with $K_{BS}=10$, $K_{RS}=6$, $M^{1}_{D}=2$ and $M^{1}_{H}=3$.}
\label{fig:markov2_new}
\end{center}
\end{figure*}

\indent For a relay based cellular OFDMA system, we have two types of incoming calls 
(as mentioned in section \ref{ICI}): 
direct and hopped calls. Let, $N_{D}$ and $N_{H}$ be the number of classes of direct calls and hopped calls respectively. 
Let, $n$ and $h$ denote the class of direct and hopped calls where  $n = 1, \ldots, N_{D} $ and  $h = 1, \ldots, 
N_{H} $. We denote the subcarrier requirement of class $n$ of direct calls and class $h$ of hopped calls by 
$M^{n}_{D}$, $M^{h}_{H BR}$ and  $M^{h}_{H RM}$ on BS-MS, BS-RS and RS-MS links respectively such that, $M^{h}_{H} = M^{h}_{H BR} + M^{h}_{H RM}$. 
Note that $M^{h}_{H}$ denotes the total subcarriers required by the class $h$ user of the hopped call. 
% Let, 
% $\mathbf{M}$ be the set of total subcarrier requirement of all classes for direct calls and hopped calls. It is represented as, 
% \begin{equation}
% \begin{split}
% \mathbf{M}= (M^{1}_{D}, M^{2}_{D}, \ldots, M^{N_{D}}_{D}, M^{1}_{HBR}, \\ M^{1}_{HRM}, M^{2}_{HBR}, M^{2}_{HRM},\ldots, M^{N_{H}}_{HBR}, M^{N_{H}}_{HRM}).
% \end{split}
% \end{equation}
If an incoming user in base region requires $M^{n}_{D}$ subcarriers, then it belongs to class $n$ of direct call 
and if it requires $M^{h}_{H}$ subcarriers in any of the relay regions, then it belongs to class $h$ of hopped call. 
Hereafter, we denote the probability of subcarrier requirement for various calls by $ \mathbb{P}_{BS-MS}\left( M_{D}^{n}\right)$, $\mathbb{P}_{BS-RS}\left( M_{HBR}^{h}\right) $ and $\mathbb{P}_{RS-MS}\left( M_{HRM}^{h}\right)$ on BS-MS, BS-RS and RS-MS links respectively. These probabilities are evaluated as illustrated in Eq.\ref{Eq_18} in Section \ref{class}. \\
\indent To admit a direct call, the required number of subcarriers should be available at the BS. However, to accomodate 
a hopped call, the required number of subcarriers should be available at BS as well as RS. Thus, a direct call 
implies the arrival of one call on BS-MS link and a hopped call implies arrival of one call each on BS-RS and RS-MS 
link\footnote{In a practical cellular system, it is ensured that RS does not receive from BS and transmit to MS simultaneously in order to eliminate the relay transmitter causing interference to its own receiver. For example, in LTE, specific subframes known as the Multicast/Broadcast Single Frequency Network (MBSFN) subframes \cite{LTE_RS_Txn} are utilized to create gaps in the RS-MS transmission, during which transmission on only BS-RS link happens. Though we have not specifically considered this scenario, our system model captures such transmission scenario if we consider resource sharing at the subframe level. Note that the analytical results remain unaffected with this consideration.} In this section, we determine the blocking probability of users belonging to direct and hopped calls.
\\ \indent We assume that call arrivals in each cell are Poisson distributed with mean arrival rate $\lambda$. Let, a fraction of the total call arrivals, say $f$ be served directly by BS, then the arrival rate of direct calls is $\lambda_{D} = f\lambda$ and that of hopped calls is $\lambda_{H} = (1-f) \lambda$. The service times of each class of direct and hopped calls are exponentially distributed with mean $\frac{1}{\mu}$. From the assumption of uniform distribution of users, hopped calls are equally distributed across the six RSs in the cell. Thus, the arrival rate of
hopped calls in the coverage area of each RS is $\lambda_{H}/6$. Let, $\mathbb{P}_{B_{D}}$ and $\mathbb{P}_{B_{H}}$ be the blocking probability of direct and hopped calls respectively. Then, the overall call blocking probability is given by,
 \begin{equation}
 \label{blocking_all}
  \mathbb{P}_{B}= f\mathbb{P}_{B_{D}}+ (1-f)\mathbb{P}_{B_{H}}.
 \end{equation}
\indent A direct call is blocked if the required number of subcarriers is not available at the BS and a hopped 
call is blocked if the required number of subcarriers is not available at any of the two i.e. BS or RS. We define 
the state of the system to be
\begin{eqnarray}
 \mathbf{S} = (M^{1}_{D}U^{1}_{D},\ldots,M^{n}_{D} U^{n}_{D}, \ldots, M^{N_D}_{D} U^{N_{D}}_{D}, M^{1}_{H}U^{1}_{H},\ldots,M^{h}_{H} U^{h}_{H},\ldots,M^{N_{H}}_{H} U^{N_{H}}_{H}),
\end{eqnarray}
where %$n=1, \ldots, N_D$ and $h=1, \ldots, N_H$ are the classes for direct and hopped calls respectively. 
$U^{n}_{D}$ and $U^{h}_{H}$ are the number of users in $n^{th}$ class of direct calls and $h^{th}$ class  of hopped calls respectively. 
% The arriving user will be blocked or not, depends upon the number of subcarriers that are in use in the system. Therefore, 
$M^{n}_{D}$ denotes the number of subcarriers required by the $n^{th}$ class user of direct call and
$M^{h}_{H}$ %implies $M^{h}_{HBR}+ M^{h}_{HRM}$, i.e., 
denotes the number of subcarriers required by the $h^{th}$ class user of hopped call.
%on BS-RS link and RS-MS link respectively.
\\ \indent These system states can be modeled by discrete time $N_{D}+N_{H}$ dimensional Markov chain. 
The state space is finite and meets the following constraints-
%the set of all possible states in the system is given by,
\begin{eqnarray}
\label{eq20}
%  \begin{split}
%  \mathcal{S} := \{\mathbf{M}.\mathbf{U} 
\sum_{n=1}^{N_D} M^{n}_{D}U^{n}_{D}+\sum_{h=1}^{N_{H}} M^{h}_{HBR}U^{h}_{H} + \sum_{h=1}^{N_{H}} M^{h}_{HRM}U^{h}_{H} \hspace{0.2in} \leq \hspace{0.2in} K_{BS}+ 6K_{RS}, 
\end{eqnarray}
\begin{eqnarray}
\label{eq21}
\sum_{n=1}^{N_D} M^{n}_{D}U^{n}_{D}+\sum_{h=1}^{N_{H}} M^{h}_{HBR}U^{h}_{H} \hspace{0.2in} \leq  \hspace{0.2in} K_{BS},  
\end{eqnarray}
\begin{eqnarray}
\label{eq22}
\sum_{h=1}^{N_{H}} M^{h}_{HRM}U^{h}_{H}\hspace{0.2in}\leq \hspace{0.2in} K_{RS}, \hspace{0.2in} \forall \hspace{0.1in}\text{RSs},
\end{eqnarray}
\begin{eqnarray}
\label{eq23}
U^{n}_{D}\hspace{0.1in}\geq \hspace{0.1in}0 \hspace{0.2in} \text{and} \hspace{0.2in} U^{h}_{H}\hspace{0.1in} \geq \hspace{0.1in} 0.
% \end{split}
\end{eqnarray}
% where,
% \begin{eqnarray}
%  \mathbf{M}.\mathbf{U} &=& \sum_{n=1}^{N_D} M^{n}_{D}U^{n}_{D} + \sum_{h=1}^{N_{H}} M^{h}_{H}U^{h}_{H}\\ \nonumber
%  &=& \sum_{n=1}^{N_D} M^{n}_{D}U^{n}_{D} + \sum_{h=1}^{N_{H}} (M^{h}_{HBR}+ M^{h}_{HRM}) U^{h}_{H}.
% \end{eqnarray}
The constraint in Eq. \ref{eq20} give an upper bound on the number of subcarriers available for allocation on the three links. The total number of subcarriers available at BS and corresponding RS gives an upper bound on the number of subcarriers that can be used in the system. The number of subcarriers available at the BS gives an upper bound on the number of subcarriers that can be used by direct calls on BS-MS link and hopped calls on BS-RS link (Eq. \ref{eq21}). Similarly, the number of subcarriers available at RS gives an upper bound on the number of subcarriers that can be used by hopped calls on RS-MS link (Eq. \ref{eq22}). In the system there can be either no user or a finite non-negative number of users on each link (Eq. \ref{eq23}).\\
 \indent \textit{Example-1}: For illustration, let us consider only one class of each call say, class 2 of direct call (i.e., $M^{2}_{D}=2$) and 
class 3 of hopped call (i.e.,$M^{3}_{H}=3$). Let $K_{BS}=10$ and $K_{RS}=6$. The number of subcarriers for hopped 
call $(M^{3}_{H}=3)$ is the sum of subcarriers required on BS-RS and RS-MS links. Note that BS-RS and RS-MS link may require either $ M^{3}_{HBR}=1$ and $ M^{3}_{HRM}=2$ or $ M^{3}_{HBR}=2$ and $ M^{3}_{HRM}=1$ depending on the SIR experienced on each link. Thus, there are two possible combinations of subcarrier requirement for a hopped call on BS-RS and RS-MS links i.e. $(1,2)$ and $(2,1)$.
Let, the probability of subcarrier requirement of hopped call be $\mathbb{P}_{H}\left(M^{3}_{H}\right)$. Then we have,
\begin{equation}
\begin{split}
\mathbb{P}_{H}\left(M^{3}_{H}=3\right)= \mathbb{P}_{BS-RS}\left(1\right)\mathbb{P}_{RS-MS}\left(2\right) +\mathbb{P}_{BS-RS}\left(2\right) \mathbb{P}_{RS-MS}\left(1\right).
\end{split}
\end{equation}
The arrival rate of hopped call of class $3$ ($\lambda^{3}_{H}$) and direct call of class $2$ ($\lambda^{2}_{D}$) is given by,
\begin{equation}
\begin{split}
\lambda^{3}_{H}= \lambda_{H}\mathbb{P}_{H}\left(M^{3}_{H}=3\right), \\
\lambda^{2}_{D}= \lambda_{D}\mathbb{P}_{BS-MS}\left(M^{2}_{D}=2\right).
\end{split}
\end{equation}
The states of the system are represented by two dimensional Markov chain in  Fig. \ref{fig:markov1}. Each state corresponds to the number of subcarrier requirement for direct calls and hopped calls. There are various combinations of different subcarrier requirement on BS-RS and RS-MS links for a hopped call. 
\\ \indent The different combinations of subcarrier requirement for the first row of Markov chain in 
Fig. \ref{fig:markov1} denotes the case, when only users of hopped call are present.  It is further illustrated in Fig. \ref{fig:markov2_new}, where the state representation is modified to indicate the number of subcarrier requirement for % direct calls on BS-MS link, 
hopped calls on BS-RS and RS-MS links distinctly. 
\\\indent 
At any instant of time, the number of calls present in the system 
using various combination of subcarriers can be found by traversing a path as shown with dotted lines in 
Fig. \ref{fig:markov2_new}. Similar combinations of subcarrier requirement of hopped calls with direct calls present in the system can be obtained for various rows of Markov chain of Fig. \ref{fig:markov1}.\\
\indent As mentioned in Example-1, when a hopped call with $M^3_H=3$ arrives in the system, it requires either of the 
combinations $(1,2)$ or $(2,1)$ subcarriers on BS-RS and RS-MS links. 
% We represent these states by $(0,1,2)$ and 
% $(0,2,1)$ respectively, where the first number indicates the number of subcarriers required for direct call, 
% which is $0$ as there are no direct calls at present. 
This hopped call is blocked when the required number of subcarriers are unavailable at either BS or RS. Observing the dotted lines in Fig. \ref{fig:markov2_new}, it becomes clear that after allocating the resources to $5^{th}$ user, BS is left with $1$ subcarrier for new allocation on BS-RS link and RS has no subcarriers left for further allocation on RS-MS link and blocking occurs. The notation $(2,1,1,1,1)$ indicates that  subcarriers $2,1,1,1,1$ are being used by different hopped calls on RS-MS link. 
Similarly, other combinations of states leading to blocking state are shown in Fig. \ref{fig:markov2_new}. This implies that there can be at most $5$ users of hopped calls in this example. \\
\indent As can be noticed from this example, determining the set of all possible states which satisfy the given constraints for a single class of 
each call is complex.
As the number of classes and the number of subcarriers at BS and RS increase, the size of the state space 
increases and it becomes very difficult to determine all possible combinations. This complexity is due to two 
reasons: a) State space consists of the subcarrier requirement of calls of all classes on all the three links and b) The states on BS-MS and BS-RS links are interrelated because BS has to use the available $K_{BS}$ subcarriers for allocation to both direct call and hopped calls.\\
\indent To simplify the computational complexity, %the state description is to be modified. Instead of defining the system state in terms of subcarriers used by direct calls and hopped calls, we define it in terms of subcarriers used at BS and RS. As BS and RS have distinct set of subcarriers, 
we consider the calls served by BS and RS as separate systems as both have distinct set of subcarriers. We also consider that for a hopped call, the required number of subcarriers are allocated by BS on BS-RS link and by RS on RS-MS link. Allocation of subcarriers to a hopped user on BS-RS and RS-MS links by BS and RS separately enables decoupling of state space of BS and RS. With this consideration, we determine the blocking probability in base and relay regions separately in the following subsections. The sum of blocking probability of calls in base region and relay region is an approximation to the overall blocking probability. We verify the validity of this approximation through simulations.
%Note that this does not affect the overall blocking probability of the system as it is the sum of the blocking probability of calls in base region and relay region. Also, the achievable rate on both the links will remain $R_{req}$ and the number of subcarriers on both links will be different based on the SIR experienced.
\\
%\indent It is to be noted that due to the orthogonal resources available at RS and BS, the resources of RS remain unutilized, when BS-MS transmission happens. This is because we do not consider any scheduling mechanism at present and focus only on `call admission' problem in this work. However, the consideration of resource partitioning between BS and RS, in both frequency and time is crucial when the problem of call admission and scheduling is to be jointly addressed.} 
\subsection{Blocking for Users present in Relay Region (Hopped Calls)}
\begin{figure*}[ht!]
\begin{center}
\scalebox{0.53}{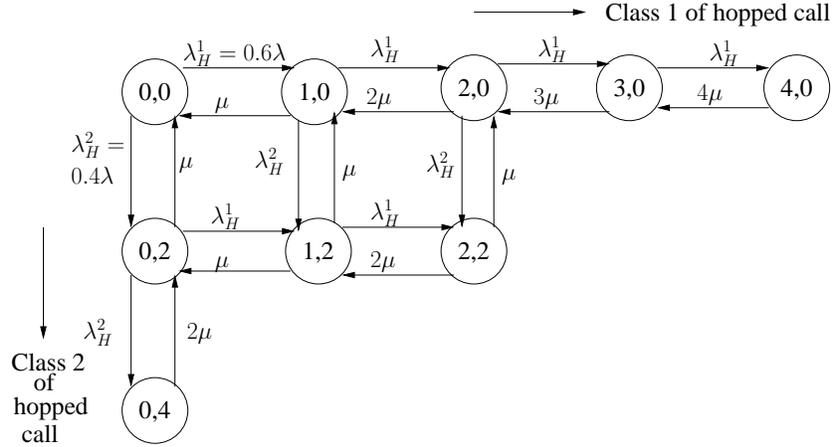}
 \caption{Two-dimensional state transition diagram  for a system with $K_{RS}=4$ and $N_{H}=2$. Incoming users are divided into 2 classes}
\label{fig:markov}
\end{center}
\end{figure*}

When a user is in any of the relay regions and experiences SIR $\gamma_{BS-RS}$ and $\gamma_{RS-MS}$ on BS-RS and 
RS-MS link, it requires $M^{h}_{HBR}$ number of subcarriers with probability $\mathbb{P}_{BS-RS}(M^{h}_{HBR})$ and 
$M^{h}_{HRM}$ number of subcarriers with probability $\mathbb{P}_{RS-MS}(M^{h}_{HRM})$. The availability of 
subcarriers on both the links i.e. BS-RS and RS-MS links is determined. If subcarriers are available on both the 
links, $M^{h}_{HBR}$ and $M^{h}_{HRM}$ subcarriers are allocated by BS and RS. Otherwise, that 
incoming user is blocked. In other words, blocking occurs when either $M^{h}_{HBR}$ number of subcarriers are unavailable on BS-RS link or
$M^{h}_{HRM}$ number of subcarriers are unavailable on RS-MS link.\\
%the number of subcarriers required by the hopped call i.e. (on both the links) is not available on either of the links. 
\indent Let, $\mathbb{P}_{B_{HBR}}$ and $\mathbb{P}_{B_{HRM}}$ be the blocking probability of hopped call on BS-RS and RS-MS link. Then, the average blocking probability of hopped calls $(\mathbb{P}_{B_{H}})$ is given as,
\begin{equation}
\label{blocking_hopped}
%\begin{split}
\mathbb{P}_{B_{H}}=1-(1- \mathbb{P}_{B_{HBR}})(1-\mathbb{P}_{B_{HRM}}).
%\end{split}
\end{equation}
\indent In this subsection, we determine $\mathbb{P}_{B_{HRM}}$ and in the next subsection we will determine $\mathbb{P}_{B_{HBR}}$.\\
There are $N_{H}$ classes of hopped calls on RS-MS link, each requiring $M^{h}_{HRM}$ subcarriers. % and share $K_{RS}$ subcarriers available at RS. 
The arrival rate of each $h^{th}$ class of these calls at RS is $\lambda^{h}_{H} = \lambda_{H} \mathbb{P}_{RS-MS}(M^{h}_{HRM})$. Let the service time for all classes of call be exponentially distributed with mean service time $\frac{1}{\mu}$. Then, the offered load for $h^{th}$ class on RS-MS link is
%$\rho^{h}_{HRM}$ = $\frac{\lambda^{h}_{H}}{\mu}$.
$\rho_{h}$ = $\frac{\lambda^{h}_{H}}{\mu}$. It is assumed that after completion of a call, the subcarriers are released by the user on both the links and they become available for use at both BS and RS.
\\ \indent We define the state of serving RS as,
\begin{eqnarray}
 \Omega_{RS} = (M^{1}_{HRM}U^{1}_H,M^{2}_{HRM}U^{2}_H,\ldots,
M^{h}_{HRM}U^{h}_H,\ldots, M^{N_H}_{HRM}U^{N_H}_H),
\end{eqnarray}
where $U^{h}_H$ is the number of users of hopped calls of 
$h^{th}$ class and $M^{h}_{HRM}$ is the number of subcarriers required by this hopped call of $h^{th}$ class.
Any class of hopped call is said to be 
blocked, when all subcarriers $K_{RS}$ are in use. Therefore, the states of the system is modeled by $N_{H}$ 
dimensional Markov chain. The state space is finite and meet the following constraints-
% the set of all possible states at the serving RS is given as,
\begin{equation}
\begin{split}
\sum_{h=1}^{N_{H}} M^{h}_{HRM}U^{h}_H \hspace{0.2in} \leq \hspace{0.2in}K_{RS}, \hspace{0.2in} U^{h}_H\hspace{0.1in}\geq \hspace{0.1in} 0 \hspace{0.2in}
%\vspace{1.5in} 
\text{ and  } \hspace{0.2in} 1 \hspace{0.1in}\leq \hspace{0.1in}h \hspace{0.1in}\leq \hspace{0.1in}N_{H}.
% \Omega_{RS} := \{\mathbf{U_{RS}}:  \mathbf{M_{RS}}.\mathbf{U_{RS}} \leq K_{RS}, U^{h}\geq 0 ,\\ 1\leq h \leq N_{H} \},
\end{split}
\end{equation}
% where,
% \begin{equation}
% \mathbf{M_{RS}}.\mathbf{U_{RS}} =  \sum_{h=1}^{N_{H}} M^{h}_{HRM}U^{h}.\\
% \end{equation}
\indent \textit{Example-2}: Let us consider $K_{RS}=4$ subcarriers and $N_{H}=2$ classes of hopped users. Let the users require $M^1_{HRM}=1$ subcarrier with probability $\mathbb{P}_{RS-MS}(M^{1}_{HRM})=0.6$ and $M^2_{HRM}=2$ 
subcarriers with probability $\mathbb{P}_{RS-MS}(M^{2}_{HRM})=0.4$. Thus, the arrival rate of 
class-$1$ users is $(\lambda_{H}^{1}) = 0.6\lambda$ and that of class-$2$ is $(\lambda_{H}^{2}) = 0.4\lambda$. 
The states of the system are denoted by $(M^{1}_{HRM}U^{1}_H,M^{2}_{HRM}U^{2}_H)$. The state transition diagram is 
shown in Fig. \ref{fig:markov}. Under the assumption of statistical equilibrium, the state probabilities are
 obtained by solving the global balance equations for each state. % using Kolmogorov's criteria.  
\\ \indent Let us consider any four interconnected states in Fig. \ref{fig:markov}. If the flow in clockwise direction equals the flow in the opposite direction, then the process is said to be reversible \cite{multi}. Let, $p(M^{1}_{HRM}U^{1}_H,M^{2}_{HRM}U^{2}_H) = p(1,2)$ be the state probability. \textbf{Note that the state probability denotes the probability that the total number of subcarriers used by class-1 and class-2 users on RS-MS link are $1$ and $2$ respectively. }Then, from Fig. \ref{fig:markov}, we have,\\
Clockwise:\\ $p(1,2) \cdot \mu \cdot p(1,0)\cdot \lambda_{H}^{1} \cdot p(2,0) \cdot \lambda_{H}^{2} \cdot p(2,2)\cdot 2\mu $.
\\Anticlockwise:\\ $p(1,2) \cdot \lambda_{H}^{1} \cdot p(2,2)\cdot \mu \cdot p(2,0) \cdot 2\mu \cdot p(1,0)\cdot \lambda_{H}^{2}$.\\
If these two expressions are equal, then the process is said to be reversible \cite{multi}.
\\ We can express any state probability, say $p(M^{1}_{HRM}U^{1}_H, M^{2}_{HRM}U^{2}_H)$ in terms of $p(0,0)$
by choosing any path between the two 
states, $p(0,0)$ and the state itself, i.e, $p(M^{1}_{HRM}U^{1}_H, M^{2}_{HRM}U^{2}_H)$ (Kolmogorov's criteria \cite{Reviewer_5}). 
\\ \indent In Fig. \ref{fig:markov}, $p(2,2)$ can be obtained by choosing the path: $(0,0)$, $(0,2)$, $(1,2)$ and $(2,2)$, and we obtain the following equation,
\begin{eqnarray}
p(2,2)&=& \frac{1}{2!} \left( \frac{\lambda_{H}^{1}}{\mu}\right)^2 \cdot \frac{\lambda_{H}^{2}}{\mu} \cdot p(0,0) \\ \nonumber
& = & \frac{(\rho_{1})^{2}}{2!} \cdot \frac{(\rho_{2})^{1}}{1!} \cdot p(0,0). 
\end{eqnarray}
Thus, there are two users of class $1$ and one user of class $2$ and this state probability has product form. Similarly, Kolmogorov's 
% This
 criteria is applicable to a system with $N_{H}$ classes, 
% the conditions for reversibility are analogue to two dimensional system and  
 %must still be fulfilled for all possible paths.
and the state probabilities in $N_{H}$ dimensional system will have product form \cite{multi}.\\
% We use this basis for finding the state probabilities in $N_{H}$ dimensional system \cite{multi}.\\
Let, $\mathbb{P}_{\Omega_{RS}}$ be the probability that the system is in state $\Omega_{RS}$. 
Since all the states are reversible, the solution is given in the standard product form \cite{multi} as,
\begin{equation}
 \mathbb{P}_{\Omega_{RS}} =  \frac{ \prod_{h=1}^{N_{H}} \frac{\rho_h^{U^h_H}}{{U^h_H}!}}{\sum_{\Omega_{RS}} \prod_{h=1}^{N_{H}} \frac{\rho_h^ {U^h_H}}{{U^h_H}!}}.
\end{equation}
Let, $\Omega_{h}$ be %the steady state of system state $\Omega_{RS}$ and thus, $\mathbf{S_{RS}}$ is 
the set of states in which an incoming hopped call on RS-MS link of either class is blocked. It is represented as,
% It is denoted by $\Omega_{h}$ and 
\begin{equation}
\Omega_{h} := \{\Omega_{h}\in \Omega_{RS} :  \sum_{h=1}^{N_{H}} M^{h}_{HRM}U^{h}_H 
%\mathbf{M_{RS}}.\mathbf{U_{RS}} 
% > K_{RS}-M^{h}_{HRM} \}.
> K_{RS} \}.
\end{equation}
\indent In Fig. \ref{fig:markov}, the states in which an incoming user of class-$1$ will be blocked are 
$\{(4,0)$, $(2,2)$ and $(0,4)\}$. The sum of the probabilities of these states is equal to the blocking probability 
for class-$1$. Similarly, the states in which an incoming user of class-$2$ will be blocked are $\{(3,0)$, $(1,2)$ 
and $(0,4)\}$. The sum of the probabilities of these states is equal to the blocking probability for class-$2$. The above illustration makes it clear that the blocking probability for any class can be obtained by summing the probabilities of all those states in which an incoming user of that class will be blocked. \\
\indent In general, for a relay based cellular OFDMA system with $N_{H}$ classes, blocking probability for hopped call of $h^{th}$ class on RS-MS link is given by,
\begin{eqnarray}
 \mathbb{P}^{h}_{B_{HRM}} & = & \sum_{\Omega_{h}}\mathbb{P}_{\Omega_{RS}} \\ \nonumber
 &=& \sum_{\Omega_h} \frac{ \prod_{h=1}^{N_{H}} \frac{\rho_h^{U^h_H}}{{U^h_H}!}}{\sum_{\Omega_{RS}} \prod_{h=1}^{N_{H}} \frac{\rho_h^ {U^h_H}}{{U^h_H}!}}.
\end{eqnarray}
The average blocking probability for hopped calls on RS-MS link is given by,
\begin{equation}
\label{blocking_RM}
\mathbb{P}_{B_{HRM}} = \sum_{h=1}^{N_{H}}\mathbb{P}^{h}_{B_{HRM}} \mathbb{P}_{RS-MS}(M^{h}_{HRM}).
\end{equation}
In the next sub-section, we determine the blocking probability of direct calls.
\subsection{Blocking for Users present in Base Region (Direct Calls)}
When a user is in base region and experiences SIR $\gamma_{BS-MS}$, it requires $M^{n}_{D}$ number of subcarriers 
with probability $\mathbb{P}_{BS-MS}(M^{n}_{D})$. The  availability of subcarriers is determined at BS. If they are 
available, then $M^{n}_{D}$ subcarriers are allocated by the BS. However $K_{BS}$ subcarriers are also shared by $h^{th}$ class of hopped call on BS-RS link, each of which requires $M^{h}_{HBR}$ subcarriers.
Thus, there are $N_{D}$ and $N_{H}$ classes of direct calls on BS-MS link and hopped calls on BS-RS link respectively. The arrival rate of $n^{th}$ class  of direct calls is $\lambda^{n}_{D}=\lambda_{D}\mathbb{P}_{BS-MS}(M^{n}_{D})$ and $h^{th}$ class of hopped calls is $\lambda^{h}_{HBR}=\lambda_{H}\mathbb{P}_{BS-RS}(M^{h}_{HBR})$.
\\ \indent We define the state of BS as, 
\begin{eqnarray}
 \Omega_{BS} = (M^{1}_{D}U^{1}_{D},M^{2}_{D}U^{2}_{D},\ldots, M^{N_{D}}_{D} U^{N_{D}}_{D}, 
M^{1}_{HBR}U^{1}_{H},M^{2}_{HBR}U^{2}_{H},\ldots,  M^{N_{H}}_{HBR} U^{N_{H}}_{H}),
\end{eqnarray}
where  $U^{n}_{D}$ is the number of direct users of $n^{th}$ class and $U^{h}_{H}$ is the number of hopped users 
of $h^{th}$ class. $M^{n}_{D}$ and $M^{h}_{HBR}$ denote the subcarrier requirement 
of $n^{th}$ and $h^{th}$ class of direct and hopped calls respectively.
If the subcarrier requirement for any class of direct call and hopped call on BS-RS link is same, then 
for BS both the calls will belong to the same class, irrespective of whether it is a direct or a hopped call. 
Thus, the state of BS can be modified as, $\Omega_{BS} = (M^{m}_{BS}U^{m}_{BS})$ where $m$ = 1,\ldots, $\max (N_{D},N_{H})$, 
denoting the class of users arriving at the BS. $M^{m}_{BS}$ denotes the subcarrier requirement of $m^{th}$ class of user 
and $U^{m}_{BS}$ denotes number of users of $m^{th}$ class arriving at the BS. \\
% \begin{equation}
% \begin{split}
% N_{D} > N_{H},     m = 1,\dots ,N_{D}  \\ 
% N_{D} < N_{H},     m = 1,\dots ,N_{H}  \\
% N_{D} = N_{H},     m = 1,\dots ,N_{D}= N_{H}  
% \end{split}
% \end{equation}
\indent It is possible that some hopped calls get the required number of subcarriers on BS-RS link but not on RS-MS link. %To consider this reduction in use of subcarriers at BS, we 
This is accounted by multiplying $\lambda_{H}$ by a discount factor $1-\mathbb{P}_{B_{HRM}}$. Let, the arrival rate of all calls at BS be $\lambda_{BS}$. Then, the arrival rate of class $m$ call at BS will be,  $\lambda_{BS}^{m}=\lambda_{D}^{m} + (1-\mathbb{P}_{B_{HRM}}^{m})\lambda_{H}^{m}$. The service time for all classes of calls at BS is exponentially distributed with mean service time $\frac{1}{\mu}$. Then, the offered load for class $m$ call at BS is $\rho_{m}$ = $\frac{\lambda_{BS}^{m}}{\mu}$.\\
\indent Any class of calls (direct or hopped calls) at BS is said to be blocked, when all 
subcarriers $K_{BS}$ are in use. Therefore, the states of the system are represented by $\max(N_{D},N_{H})$ dimensional Markov chain. The state space is finite and the constraints to be met are,
\begin{equation}
\begin{split}
\sum_{m=1}^{\max(N_D,N_H)} M^{m}_{BS}U^{m}_{BS}
% \Omega_{BS} := \{\mathbf{S_{BS}}:  \mathbf{M_{BS}}.\mathbf{U_{BS}} 
\hspace{0.1in}\leq \hspace{0.1in}K_{BS}, \hspace{0.1in}U^{m}_{BS}\hspace{0.1in}\geq \hspace{0.1in}0 \ ,  \hspace{0.2in} 1 \leq m \leq \max(N_{D},N_{H}).
\end{split}
\end{equation}
\indent Let, $\mathbb{P}_{\Omega_{BS}}$ be the probability that the system is in state $\Omega_{BS}$. Since all the states are reversible, the solution is given in the product form as following \cite{multi},
\begin{equation}
\mathbb{P}_{\Omega_{BS}} =  \frac{\prod_{m=1}^{\max(N_{D},N_{H})}\frac{(\rho_{m})^{U^{m}_{BS}}}{U^{m}_{BS}!}}{\sum_{\Omega_{BS}}\prod_{m=1}^{\max(N_{D},N_{H})}\frac{(\rho_{m})^{U^{m}_{BS}}}{U^{m}_{BS}!}}.
\end{equation} 
\indent Let, $\Omega_{m}$ be the %steady state of system state $\Omega_{BS}$ and thus $\mathbf{S_{BS}}$ is the 
set of those states in which an incoming direct call or hopped call at BS of any class is blocked.
It is represented as,
% denoted by $\Omega_m$ and is 
\begin{equation}
%\begin{split}
\Omega_{m} := \{\Omega_{m} \in \Omega_{BS} :  \sum_{m=1}^{\max(N_D,N_H)} M^{m}_{BS}U^{m}_{BS} > K_{BS}\}.
% \Omega_{m} := \{\Omega_{m} \in \Omega_{BS} :  \sum_{m=1}^{\max(N_D,N_H)} M^{m}_{BS}U^{m}_{BS} > K_{BS}-M^{m}_{BS} \}.
%\end{split}
\end{equation}
Therefore, the blocking probability for $m^{th}$ class user at BS is given by,
\begin{eqnarray}
 \mathbb{P}^{m}_{B_{BS}} & = & \sum_{\Omega_{m}}\mathbb{P}_{\Omega_{BS}} \\ \nonumber
 &=& \sum_{\Omega_m}\frac{ \prod_{m=1}^{\max(N_{D},N_{H})} \frac{(\rho_{m})^{U^{m}_{BS}}}{U^{m}_{BS}!}}{\sum_{\Omega_{BS}} \prod_{m=1}^{\max(N_{D},N_{H})} \frac{(\rho_{m})^{U^{m}_{BS}}}{U^{m}_{BS}!}}.
\end{eqnarray}
The average blocking probability for direct calls on BS-MS link $(\mathbb{P}_{B_{D}})$ and hopped calls on BS-RS link $(\mathbb{P}_{B_{HBR}})$ is given by,\\
\begin{equation}
\label{blocking_BS}
\begin{split}
\mathbb{P}_{B_{D}} = \sum_{m=1}^{N_{D}}\mathbb{P}^{m}_{B_{BS}} \mathbb{P}_{BS-MS}(M^{m}_{D}), \\
\mathbb{P}_{B_{HBR}} = \sum_{m=1}^{N_{H}}\mathbb{P}^{m}_{B_{BS}} \mathbb{P}_{BS-RS}(M^{m}_{HBR}).
\end{split}
\end {equation}
Thus, from Eqs. \ref{blocking_all}, \ref{blocking_hopped}, \ref{blocking_RM} and \ref{blocking_BS}, we can determine the overall blocking probability of the system.
\section{Results and Discussions}
\label{numericalresults}
\subsection{Comparison of Analytical and Simulation Results}
\label{numericalresults1}
\begin{figure}[h!]
 \begin{center}
 \includegraphics[width=1\textwidth, height= 3.6in]{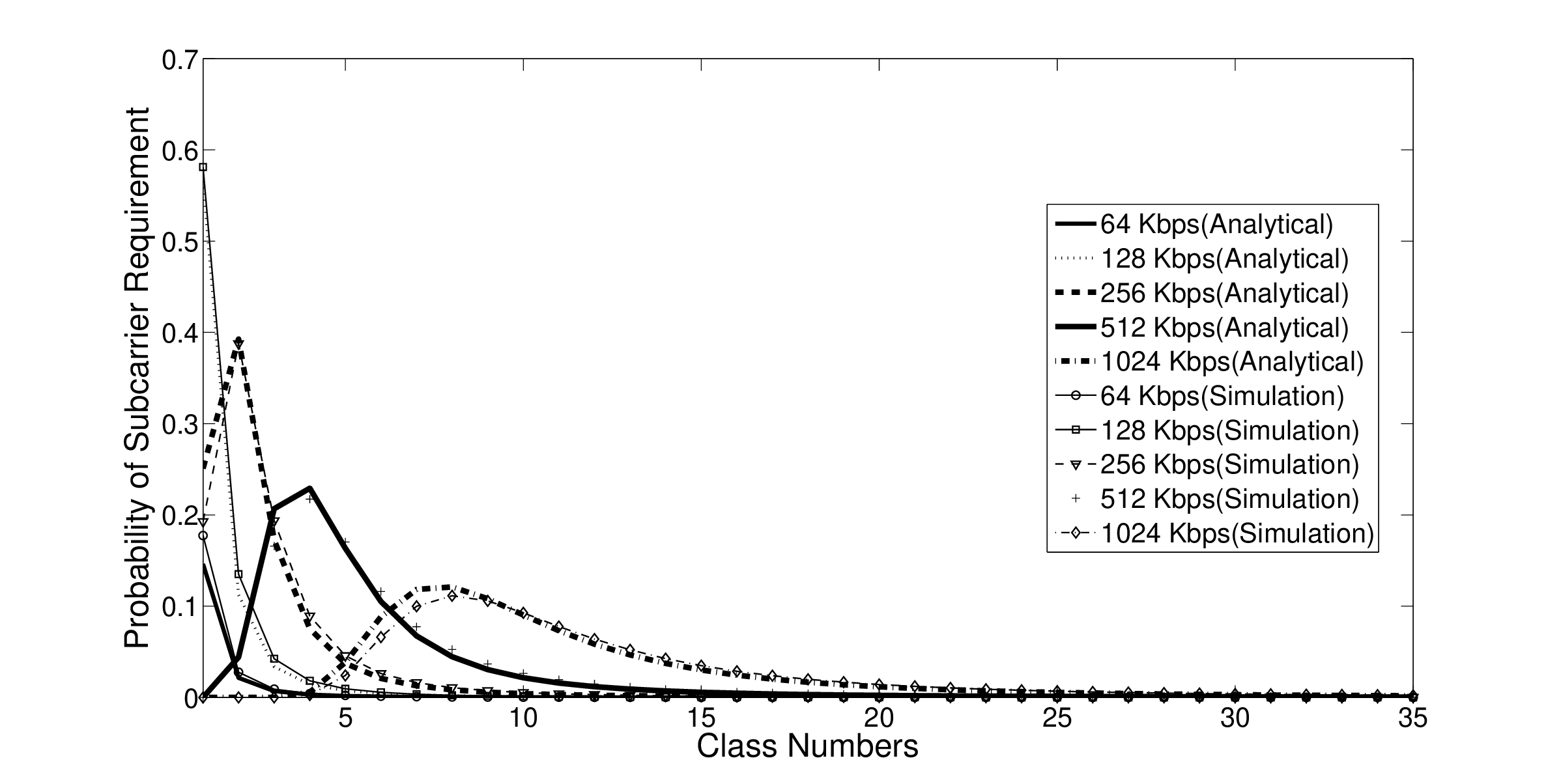}
\caption{Probability of Subcarrier requirement (i.e. class) for varying rate requirements on BS-MS link (Subcarrier BW = $15$ KHz)}
 \label{Fig1}
\end{center}
 \end{figure}

\begin{figure}[h!]
 \centering
 %\scalebox{0.4}
 %\includegraphics [scale=0.49] {Fig2.eps}
 \includegraphics[width=1\textwidth, height= 3.6in]{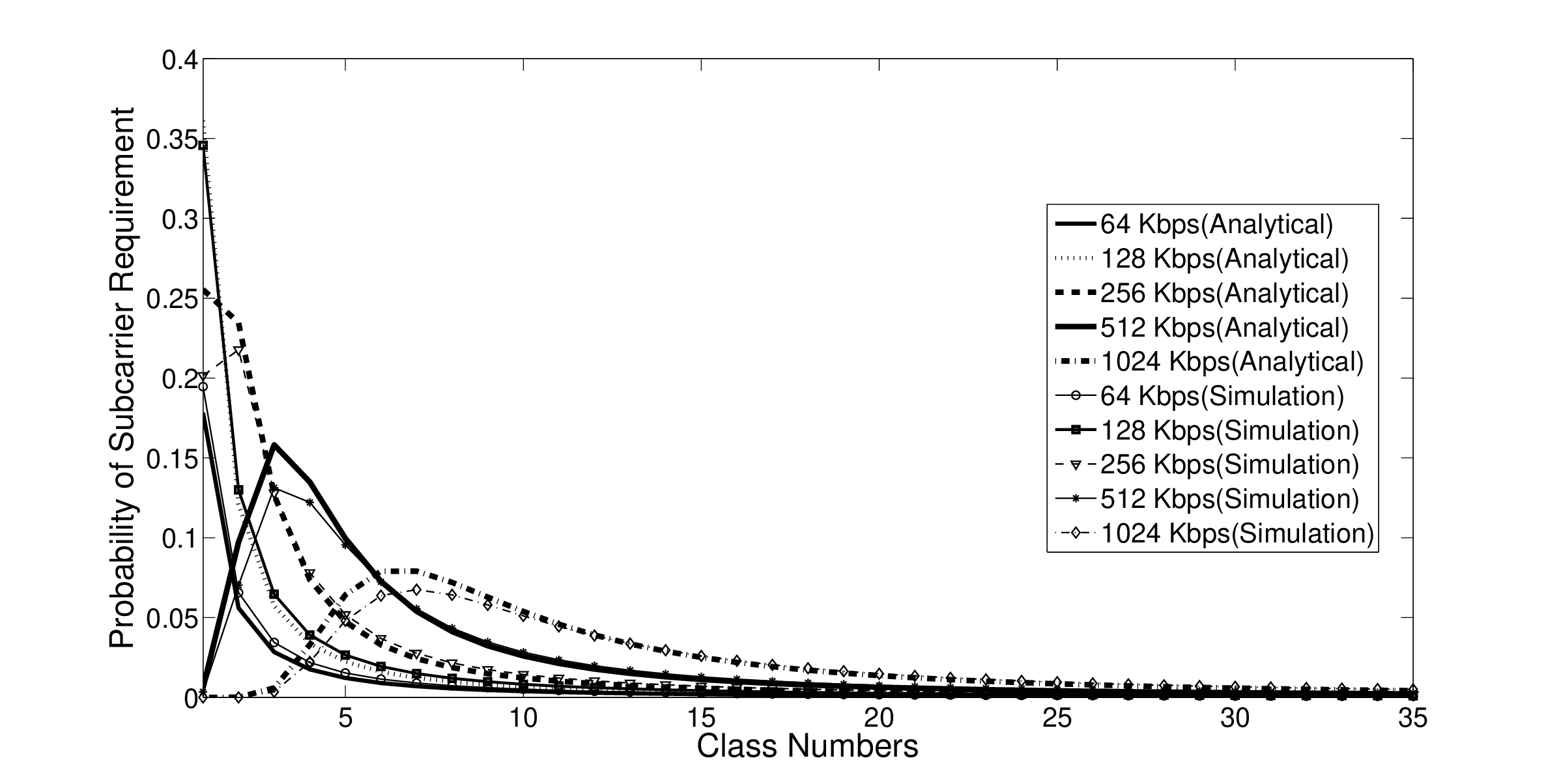}
 \caption{Probability of Subcarrier requirement (i.e. class) for varying rate requirements on BS-RS link (Subcarrier BW = $15$ KHz)}
 \label{Fig2}
 \end{figure}

\begin{figure}[h!]
 \begin{center}
  \includegraphics[width=1\textwidth, height= 3.6in]{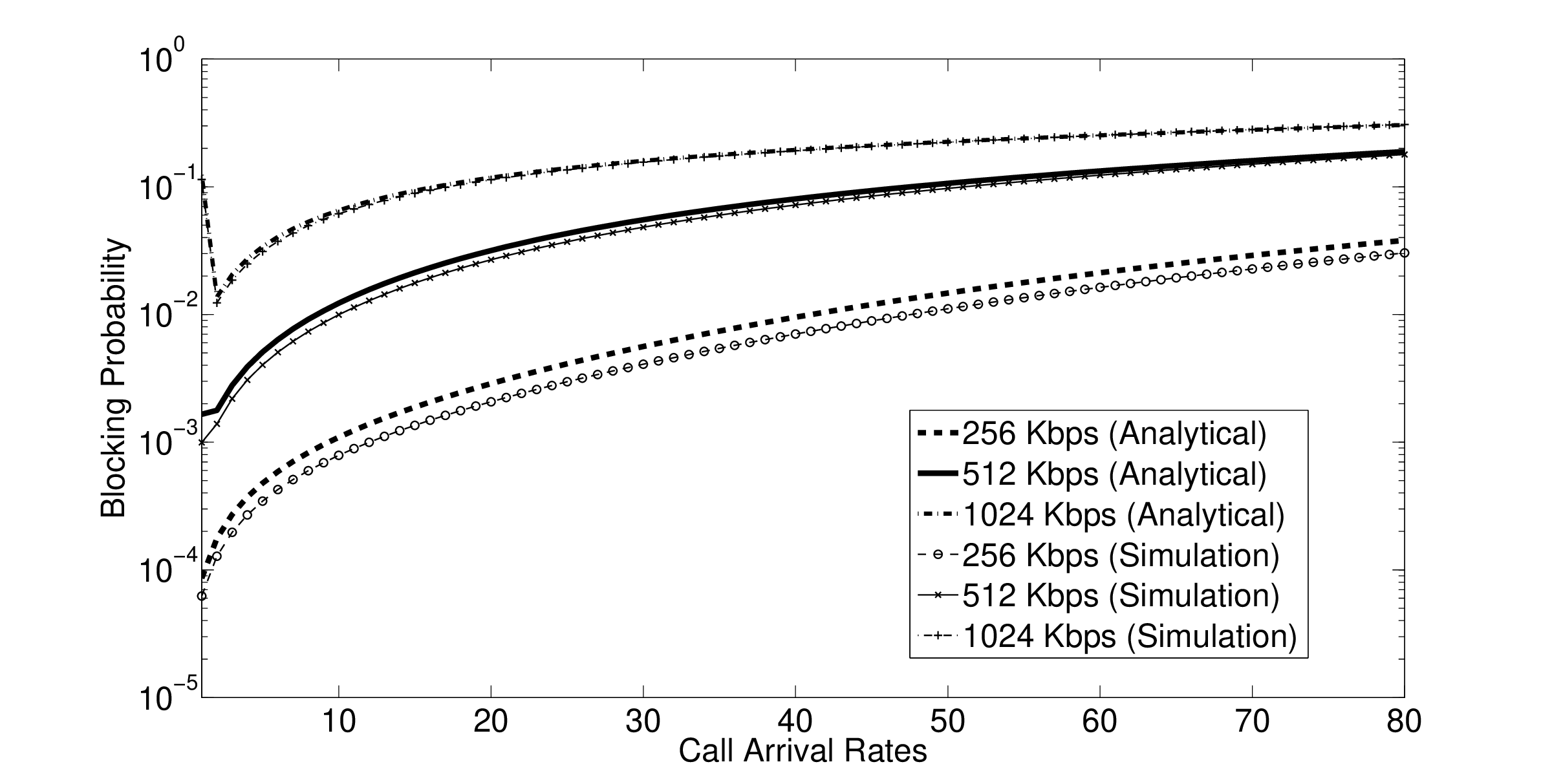}
 \caption{Impact of Rate Requirement on Blocking Probability  when Subcarrier BW = $15$ KHz (Number of classes for 256 Kbps = $15$, Number of classes for $512$ and $1024$ Kbps = $31$)}
 \label{blocking_arrival1}
\end{center}
 \end{figure}

\begin{figure}[h!]
 \centering
 %\scalebox{0.4}
\includegraphics[width=1\textwidth, height= 3.6in]{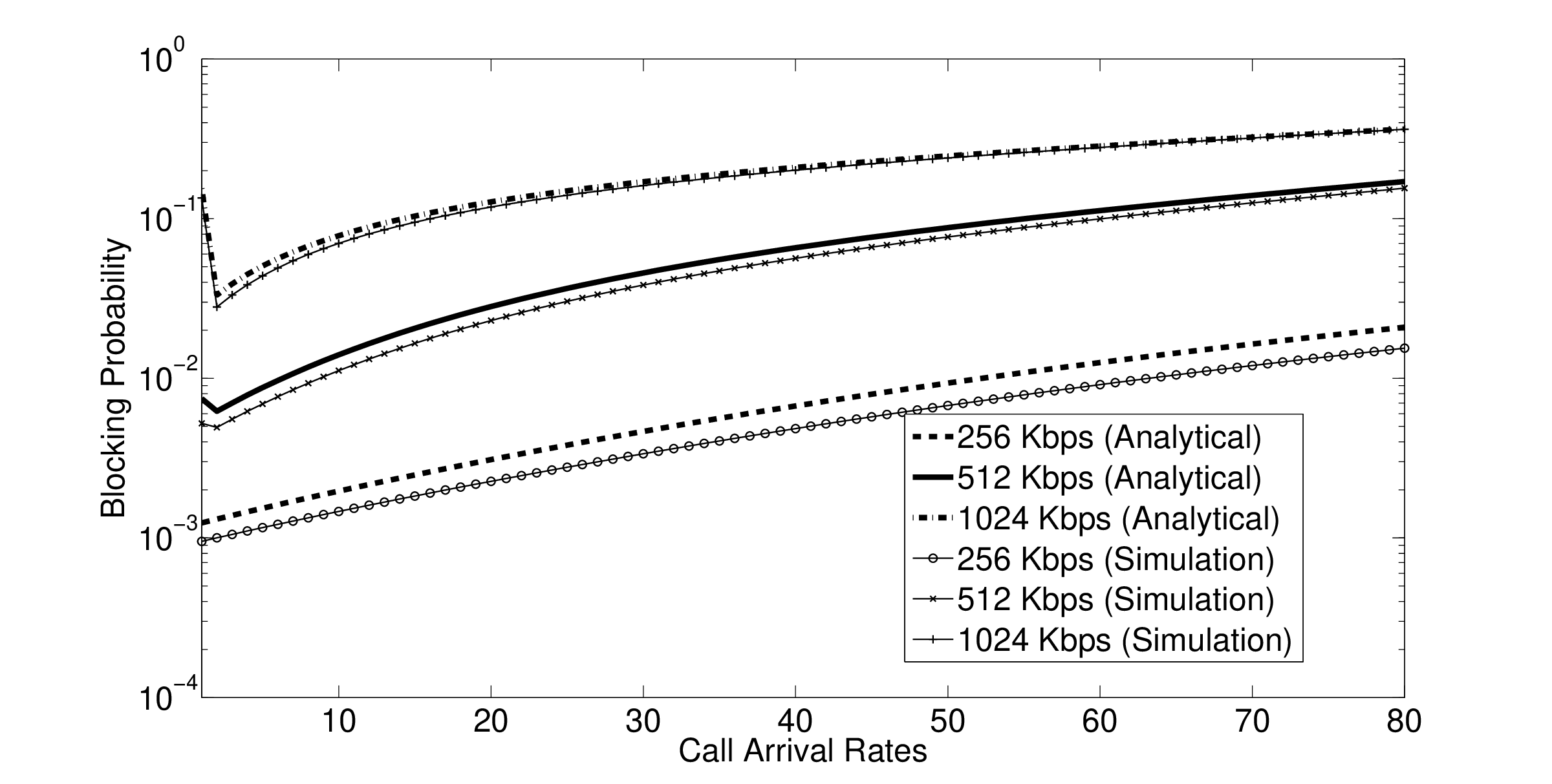} 
\caption{Impact of Rate Requirement on Blocking Probability when Subcarrier BW = $30$ KHz (Number of classes for 256 Kbps = $15$, Number of classes for $512$ and $1024$ Kbps = $31$)}
 \label{blocking_arrival2}
 \end{figure}

\begin{figure}[h!]
 \centering
 %\scalebox{0.4}
 %\includegraphics [scale=0.5] {Fig5.eps}
 \includegraphics[width=0.9\textwidth, height= 3.6in]{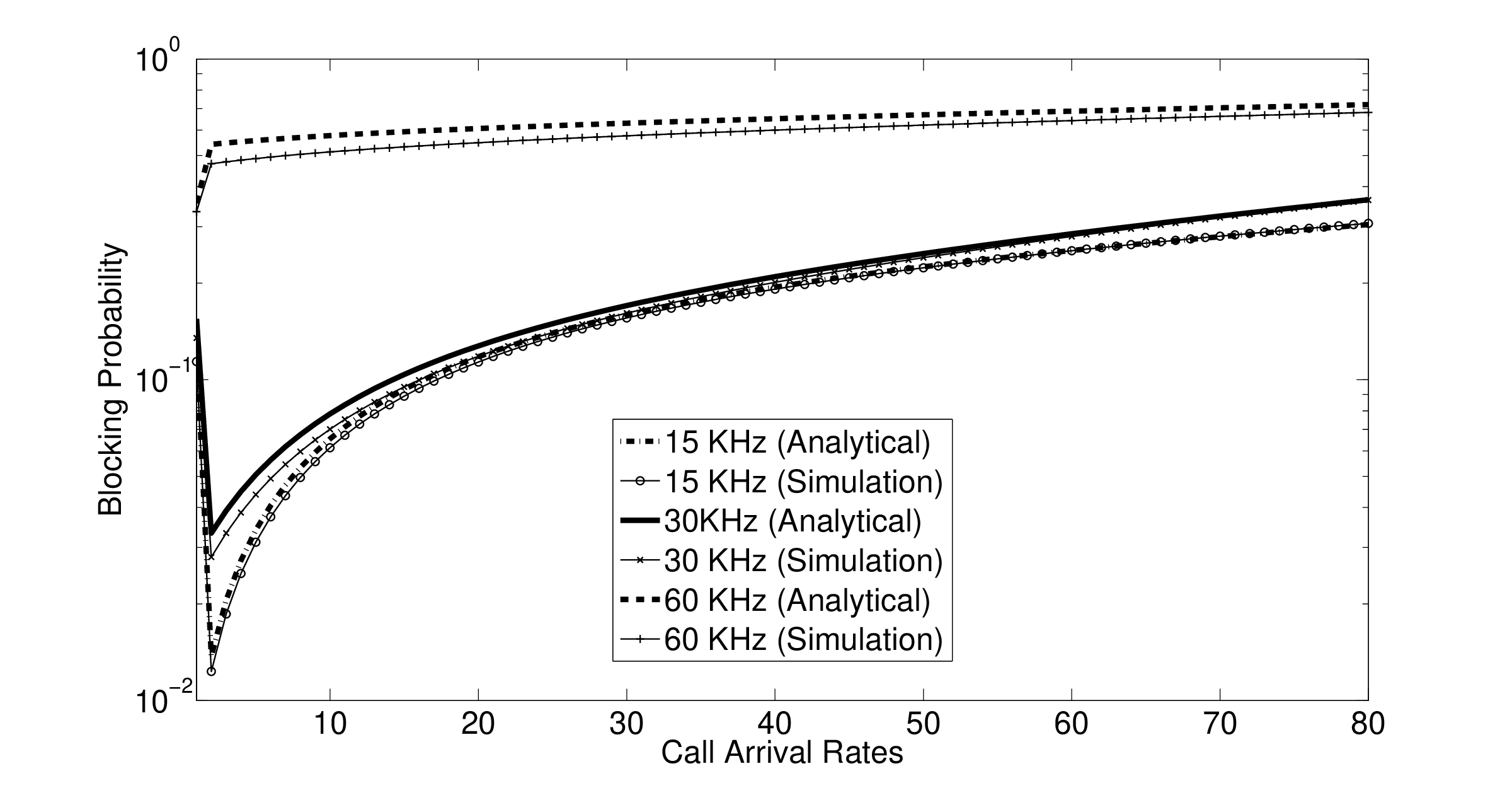}
  \caption{Variation in Blocking Probability with change in Subcarrier BW for a given rate requirement = $1024$ Kbps and Number of classes = $31$}
 \label{Fig5}
 \end{figure}

In this section, we illustrate the results based on the analytical models developed in the previous sections and present validation of the analytical results using simulations. We consider
the downlink of relay assisted OFDMA system. The values of system parameters chosen for the analysis are as per the LTE standard \cite{3GPP_PHY} and are given in Table \ref{simulation}. 
% Effect of Subcarrier BW
We perform the analysis considering four rate requirement - $64, 128, 256$ and $1024$ Kbps.\\\indent
The simulation procedure consists of modeling a snapshot of location of users (calls), their arrival and departure times in the reference cell as well as neighboring cells. The user can be located either in the base region or relay region of a cell. We generate a fraction $(f)$ of total calls in the base region and remaining in the relay region. The call arrivals are Poisson distributed with rate $\lambda$ and holding times are exponentially distributed with mean $\frac{1}{\mu}$ in all cells. Available subcarriers $K$ are shared between BS and six RSs.\\ 
% % %Each RS and BS are allocated $K_{RS}$ and $K_{BS}$ subcarriers respectively. We consider inter-cell interference from BSs of the first-tier of neighboringcells only.\\ 
\indent For every new call arrival, we check the association of user with base region or relay region. Based on this association, a call is termed as direct call or hopped call. Accordingly, we evaluate the SIR experienced by that call on BS-MS link (or BS-RS and RS-MS links). We consider the random subcarrier allocation scheme on all the three transmission links. For a direct (hopped) call on BS-MS link (BS-RS and RS-MS links), one subcarrier is randomly chosen from the available subcarriers, i.e., the unused subcarriers from the total of $K_{BS}$ for BS and $K_{RS}$ for RS. Then, it is checked whether the user's rate requirement is satisfied, that is whether $\log_{2}(1+\gamma_{BS-MS})$ for that subcarrier is greater than or equal to required rate $(R_{req})$. If not, BS or RS continues to add randomly chosen subcarriers until the total achievable rate become greater or equal to $R_{req}$. If the available set of subcarriers can not meet the rate requirement, the call is blocked. Note that a hopped call is blocked if the required number of subcarriers are not available on either of the links. We consider that the set of allocated subcarriers to the user is utilized for the entire duration of the  call. After the completion of call, the subcarriers are released by the user and they become available for use simultaneously at both BS and RS. At this point, the processing of one snapshot is complete and another snapshot is continued. Simulation is performed over such $10,000$ independent snapshots. From these simulations, we determine the probability distribution of the subcarrier requirement on each of the links. 
\\
\indent Fig. \ref{Fig1} and Fig. \ref{Fig2} give the probability of subcarrier requirement (evaluated in Eq. \ref{Eq_18}) for four different data rates on BS-MS and BS-RS link respectively. It is the probability of a call belonging to a certain class. The probability of a call belonging to lower class is more on BS-RS link due to line of sight path and lesser impact of shadowing.
From these two figures, we observe that the subcarrier requirement changes with the change in rate requirement ($R$). For lower $R$, less number of subcarriers are required and therefore, majority of users will belong to lower classes. For example, in Fig. \ref{Fig1}, for $R=64$ Kbps, the probability of a call belonging to a class between 1 and 15 is non-zero, and the probability of a call belonging to higher classes is close to zero. Similarly, for high $R$, say $R=1024$ kbps, there are effectively no users that require less number of subcarriers. Therefore, the probability of a call belonging to a class between 5 and 35 is non-zero, and the probability of call belonging to lower classes is almost zero. This clearly indicates two things-
that for a given rate requirement, 1) the number of class will depend on the range for which the probability of a call belonging to a certain class (i.e., 
subcarrier requirement) is non-zero and 2) the definition of class will not always be $M^r=r$ (as defined in Section \ref{class}), in particular for higher rates. It will depend on the lowest and the highest class for which the subcarrier requirement is non-zero. For example, for $R=1024$ Kbps, there will be $31$ classes and it will range from class $5$ to class $35$, i.e., $M^r= r + 4$ (offset of 4).
Thus, definition of class and determining the number of classes depend on the rate requirement. We observe that the simulation results closely match with the analytical results.
\\\indent Fig. \ref{blocking_arrival1} and Fig. \ref{blocking_arrival2} illustrate the impact of rate requirement on blocking probability (evaluated in Eq. \ref{blocking_all}) for two cases: when the subcarrier bandwidth considered is of $15$ KHz and $30$ KHz. We observe that increasing the subcarrier 
bandwidth results in an increase in the blocking probability. It is because when the subcarrier bandwidth is more, 
the total number of subcarriers available in the system reduces. In this case, even though a user may meet
its rate requirement with fewer number of subcarriers, the overall blocking probability is likely to increase. Similar observation can be made from Fig. \ref{Fig5} where the blocking probability is computed for three different subcarrier bandwidth, $15$, $30$ and $60$ KHz, for fixed rate requirement of $1024$ Kbps.
% Effect of Rate
\\\indent From Fig. \ref{blocking_arrival1} and Fig. \ref{blocking_arrival2}, we also observe 
%, Fig. \ref{blocking_arrival2}, Fig.\ref{blocking_arrival3} 
% and Fig. \ref{blocking_arrival4} 
that irrespective of the subcarrier bandwidth, the blocking probability is influenced by the rate requirement of users.
As the rate requirement of users increase, they will
require more number of subcarriers and therefore, blocking probability increases. For the simulations, we count the number of times an incoming call is blocked and plot the blocking probability of the system. We observe a good agreement between analytical and simulation results.\\
%%%Effect of class
%%Note that the class is an indicative of the number of subcarriers required by the user.
%%Therefore, more diverse the subcarrier requirement of the users, higher will be the blocking probability.
%%Increase in blocking probability with the increase in the number of classes is illustrated again with more clarity in Fig.
%%\ref{blocking_arrival5}.\\
%
\indent As an intuitive insight, when the subcarrier bandwidth is high, 
the blocking probability is influenced by the number of users belonging to lower classes and higher classes.
If majority of users belong to lower classes, then the subcarrier bandwidth will result in an allocation which will be much more 
than their requirement, leading to an inefficient resource utilization and hence, increase in blocking probability. 
On the other hand, if majority of users belong to higher classes, 
then they will quench their resource requirement in fewer resources and blocking probability is likely to reduce.
However, in general when the probability that a user belongs to a higher or a lower class (i.e., the subcarrier requirement
of a class is more or less) is equal, 
an increase in the subcarrier bandwidth will reduce the 
number of resources available in the system and hence there will be an increase in the blocking probability.\\
\subsection{Comparison of Non-Relay System with Relay-based OFDMA System through simulations}
\begin{figure}[h!]
   \centering
   \subfigure[Blocking Probability for non-relay Cellular OFDMA System]
   {%[width=1\textwidth, height= 4.4 in]
    \includegraphics[scale=0.85]{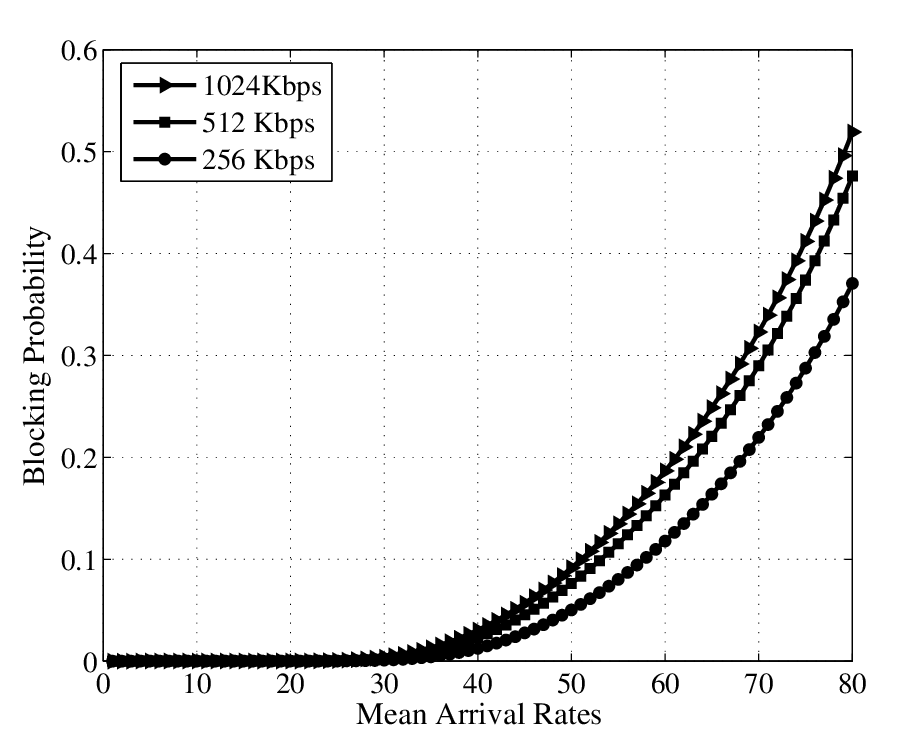}
    \label{blocking_ofdma1}
    }
   \subfigure[Blocking Probability for Relay based Cellular OFDMA System]
   {
   \includegraphics[scale=0.85]{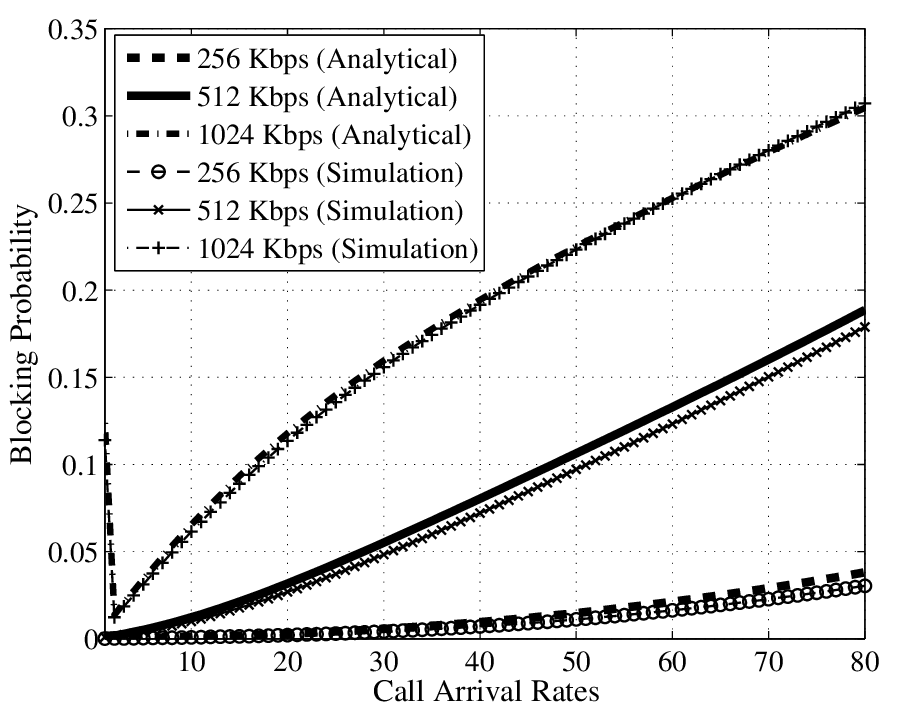}
   \label{blocking_relay1}
   }
   \caption{Blocking Probability versus Mean Arrival Rate for Various Rate Requirements a) without Relays and b) with Relays in a cellular OFDMA system}
   \label{fig:blocking_relay_ofdma}
\end{figure}

\indent Figure \ref{blocking_ofdma1} and \ref{blocking_relay1} illustrate the impact of rate requirement on blocking probability for two cases: cellular OFDMA system without and with relays for subcarrier bandwidth of $15$ KHz. We observe that the blocking probability in relay based cellular OFDMA system is much lower than the system without relays. We observe that in relay based cellular OFDMA systems, for higher rate requirements such as, $1024$ Kbps, the blocking probability is reduced by only $10\%$. However, for lower rate requirements, such as $512$ Kbps, the blocking probability reduces by about $50\%$. This demonstrates that relay deployment decreases the blocking probability and hence improves the capacity. We also observe that %there is a perfect match in the simulation and analytical results in case of non-relay OFDMA system, while 
there is a close match in the simulation and analytical results in case of relay-based OFDMA system.
\\\indent \textbf{
We can also observe the impact of arrival rate on the blocking probability in both cases. %For the ease of discussion, we divide the arrival rates into two regimes, lower and higher. 
For lower arrival rate scenario, the blocking probability is higher for the system with relays compared to the system without relays. This happens because a hopped call is blocked when the required number of subcarriers are unavailable on either of the two links, BS-RS and RS-MS. With lower arrival rate, the possibility of having users of all classes is also less. For instance, if there are more users belonging to higher classes, then the probability of hopped call blocking may increase, thereby increasing the overall blocking probability of the system. For higher arrival rate scenario, the distribution of users belonging to different classes is likely to be more uniform. This %leads to better utilization of resources by 
results in admitting lower class users if the resources are insufficient for a higher class user and thus, blocking probability reduces in a system with relays, for higher arrival rate scenario. Thus, we can infer that a system with relays offers significant reduction in blocking probability compared to a system without relays, for higher arrival rate scenario. }
% However, the discrepancy in the percentage improvement for higher and lower rate requirement cases is due to the following reason.
% $512$  and $1024$ Kbps, the blocking probability is reduced by $50\%$ and $10\%$ respectively. However, for lower rate requirements, such as $256$ Kbps the blocking probability reduces substantially. This demonstrates that relay deployment decreases the blocking probability and hence improves the capacity. 
% 
% However, in a scenario where large number of subcarriers are required at RSs, blocking may increase at RS. Hence overall blocking probability of the system may degrade and no benefit of deploying RSs could be achieved. For example, for higher rate requirements ($1024$ Kbps), blocking probability with RSs is reduced by just $10\%$ in comparison to lower rate requirements where substantial reduction in blocking probability is observed. It happens because we have partitioned the total available subcarriers $(K)$ between BS and RS heuristically. This discrepancy can be resolved and blocking probability can be reduced further with optimal resource partitioning between BS and RSs.
\\
 \indent 
Note that due to delay tolerant characteristic, data calls can be queued (delayed) and can be analyzed in terms of waiting time probability, i.e., the probability that a queued user gets service within the maximum acceptable waiting time. Thus, blocking probability analysis using \textit{Erlang loss model} can not be applied for data calls.
Following standard queuing theory, it is known that very small blocking probability in \textit{Erlang loss model} can also achieve small delay in equivalent \textit{Erlang delay model}. It is due to this fact, analysis for voice calls at higher rates is also applicable for data calls of higher rate services such as, video downloads, video streaming, multimedia conferencing, on-line gaming etc.
In general, blocking probability for voice traffic $(\mathbb{P}_B)$ and waiting time probability for data traffic ($\mathbb{P_D}$) are related as follows, 
\begin{equation}
\mathbb{P}_D = \mathbb{P}_B \frac{K}{K-\rho}, \hspace{0.2 in} K > \rho
\end{equation} 
where, $K$ is the number of resources (subcarriers) available in the system and $\rho$ is the offered traffic.  
This implies that the waiting time probability is greater than blocking probability by a factor of $\frac{K}{K-\rho}$. The determination of 
blocking probability of voice calls for higher data rates may be helpful in the determination of waiting time probability of data calls.
Therefore, we have performed the blocking probability analysis of voice calls for rate requirements as $256$, $512$ and $1024$ Kbps.
 \section{Conclusions and Future Work}
\label{conclusion}
In cellular OFDMA networks, in order to meet the same rate requirement, the number of subcarriers 
required are different for different users (due to differences in their locations and experienced SIR) on various 
links. Therefore traditional methods of blocking probability computation cannot be used directly. We have
proposed an analytical model to evaluate blocking probability for relay based cellular OFDMA networks. The CDF of SIR is
determined to compute the probability distribution of subcarriers required on 
the three transmission links. The incoming users are classified into different classes based on their 
subcarrier requirement. We have modelled such a system by a multi-dimensional Markov chain.
%  and observed that the complexity increases due to large state space when both BS and RS system states are considered. To reduce 
% this complexity, we decouple the state space of BS and RS. This assumption gives an upper bound on 
% the overall blocking probability, which is the sum of blocking probabilities of calls at BS and RS.
The effects of  subcarrier bandwidth ($W$) and rate requirement ($R$) on the blocking probability are analyzed. We have also analyzed the effect of rate requirement on the definition of class and number of classes. %This model is useful in the performance evaluation, design, planning of resources and call admission control of relay based cellular OFDMA networks like LTE. 
\\\indent 
We have considered six relays per cell in our system model, with their locations fixed in the cell. %determined heuristically. 
%This model can be used to study the impact of number of RSs. 
However, the optimal location of RSs can impact the system performance in terms of improving cellular coverage or network capacity. One of the author's work \cite{NCC_Mahima} considers the optimal relay placement problem in the context of maximizing the cellular coverage. The optimal relay positioning to maximize capacity and reduce blocking probability is can be investigated as future work. \\ %will be considered in future.\\  
\indent Though we do not consider multi-service traffic where each class of users has a different rate requirement, our analysis can be extended to such a scenario.
We have considered uniform distribution of users and partitioned total available subcarriers $(K)$ between BS and RS in somewhat heuristic manner. However practically, there may be non-uniform traffic in relay region and an optimal resource partitioning scheme can be designed.% We are currently investigating the impact of 1) number of RSs, 2) optimal location of RSs and 3) flexible and optimal resource partitioning between BS and RSs on the blocking probability.}\\ 
% \textbf{
% \indent It is to be noted that due to the orthogonal resources available at RS and BS, the resources in half the time slots at RS remain unutilized, when BS-MS transmission happens. This is because we do not consider any scheduling mechanism at present and focus only on `call admission' problem in this work. However, 
% %the consideration of 
% an optimal resource partitioning between BS and RS can be beneficial  %, in both frequency and time is crucial 
% when the problem of call admission and scheduling is to be jointly addressed.}
% \\\indent
\section{Acknowledgement}
This  work  is  supported by the India-UK
Advanced Technology of Centre of Excellence in Next Generation Networks (IU-ATC)  project  and  funded  by  the Department of Science and
Technology (DST), Government of India.
\bibliographystyle{ieeetr}
\bibliography{wcnc1_ref.bib}
% ======================================================
%----- NOTION TABLE------------------------
\begin{table}[!t]
\caption{List of notations used in Section IV and V}
\begin{tabular}{c l}
\hline
Symbol & Description\\
\hline
\hline
\\
$\gamma^{m}_{BS-MS}$ & SIR experienced by a user while using $m^{th}$ subcarrier on BS-MS link. \\
$ R$ & Rate requirement of incoming users\\
$\mathcal{M}$ & Number of subcarriers required by any user \\
$ W$ & Subcarrier Bandwidth \\
$I_{BS-MS}$ & Interference to signal ratio on BS-MS link\\
$l$ & Number of non-overlapping consecutive intervals into which the $I_{BS-MS}$ is divided\\
$r$ & Class of a user \\
$M^r$ & Number of subcarriers required by a user of class $r$\\
$\mathbb{P}_{BS-MS}(M^r) $ & Probability that an incoming user on a direct call belongs to class $r$\\
$\mathbb{P}_{H}(M^r) $ & Probability that an incoming user on a hopped call belongs to class $r$\\
$\mathbb{P}_{BS-RS}(M^r) $ & Probability that an incoming user on BS-RS link belongs to class $r$\\
$\mathbb{P}_{RS-MS}(M^r) $ & Probability that an incoming user on RS-MS link belongs to class $r$\\
$F_{\mathbf{I_{BS-MS}}}\left( I^{r}_{BS-MS}\right) $ & CDF of interference to signal ratio on BS-MS link\\
$N_D$ & Number of classes of direct calls\\
$N_H$ & Number of classes of hopped calls\\
$M^n_D$ & Number of subcarriers required by $n^{th}$ class user of direct call\\
$M^h_H$ & Total number of subcarriers required by $h^{th}$ class user of hopped call on both links\\
$M^h_{HBR}$ & Number of subcarriers required by $h^{th}$ class user of hopped call on BS-RS link\\
$M^h_{HRM}$ & Number of subcarriers required by $h^{th}$ class user of hopped call on RS-MS link\\
$M^m_{BS}$ & Number of subcarriers required by $m^{th}$ class user of direct or hopped call, served by BS\\
$\lambda$ & Mean arrival rate of each call \\
$\rho_h$ & Offered load for $h^{th}$ class of hopped call\\
$f$ & Fraction of calls served directly by BS\\
$\lambda_D$ & Arrival rate of direct calls\\
$\lambda_H$ & Arrival rate of hopped calls\\
$\lambda^h_H$ & Arrival rate of $h^{th}$ class of hopped calls\\
$\mathbb{P}_{B_D}$ & Blocking probability of direct call\\
$\mathbb{P}^m_{B_{BS}}$ & Blocking probability of $m^{th}$ class user at BS\\
$\mathbb{P}_{B_H}$ & Blocking probability of hopped call\\
$\mathbb{P}_{B_{HBR}}$ & Blocking probability of hopped call on BS-RS link\\
$\mathbb{P}_{B_{HRM}}$ & Blocking probability of hopped call on RS-MS link\\
$\mathbb{P}_{B}$ & Overall blocking probability\\
$S$ & State of the system\\
$U^n_D$ & Number of users in $n^{th}$ class of direct calls\\
$U^h_H$ & Number of users in $h^{th}$ class of hopped calls\\
$U^m_{BS}$ & Number of users in $m^{th}$ class of direct or hopped call, served by BS\\
$\Omega_{RS}$ & State of serving RS \\
$\Omega_{BS}$ & State of BS\\
$\Omega_{h}$ & Set of states in which an incoming hopped call on RS-MS link is blocked\\
$\Omega_{m}$ & Set of states in which an incoming direct or hopped call at BS is blocked\\
$\mathbb{P}_{\Omega_{RS}}$ & Probability that the system is in state $\Omega_{RS}$\\
$\mathbb{P}_{\Omega_{BS}}$ & Probability that the system is in state $\Omega_{BS}$\\

\\

\hline
\end{tabular}
\label{symbols}
\end{table}
% =======================================================
   \begin{table}[h!]
 \centering
 \caption{Parameters for Numerical Analysis}
 \begin{tabular}{|l|l|}
 \hline
%  \textbf{Simulation Parameters} & \textbf{Value} \\
\textbf{Parameters for Numerical Analysis} & \textbf{Value} \\
\hline Inter BS distance (meters) & $1732$ \\
 \hline System Bandwidth (MHz) & 10  \\
 \hline Subcarrier bandwidth $(W)$ (KHz) & $15$ ($30$) \\
 \hline Number of Subcarriers available at & $480$ (when\\
 BS $(K_{BS})$ &  $W = 15$ KHz)\\
\ & $240$ (when \\ & $W = 30$ KHz) \\
 \hline Number of RSs in each reference cell & $6$\\
 \hline Number of subcarriers available at & $30$ (when\\
 each RS $(K_{RS})$ & $W = 15$ KHz)\\ & $15$ (when\\ & $W = 30$ KHz) \\
 \hline Number of interferring cells from first  & \\
 tier $(N)$ & $6$ \\
 \hline Path loss exponent $(\beta)$ & 3.5 \\
 \hline Shadowing standard deviation on &\\
BS-MS link $(\sigma_{BS-MS}=\sigma_{iBS-MS})$ & $8$ dB\\
and RS-MS link $(\sigma_{RS-MS}=\sigma_{iRS-MS})$ & \\
% \hline Shadowing standard deviation on &\\
%  RS-MS link $(\sigma_{RS-MS}=\sigma_{iRS-MS})$ & $6$ dB\\ 
\hline Shadowing standard deviation on &\\
  BS-RS link $(\sigma_{BS-RS}=\sigma_{iBS-RS})$ & $4$ dB\\
 \hline Rate requirement of each call $(R)$ (Kbps)& $64, 256,$ \\ & $512, 1024$  \\
 \hline Maximum number of classes for each call & \\
 $(N_{D}$ = $N_{H})$ & $10$ ($15$)\\
\hline Fraction of calls arriving at BS $(f)$ & 0.5\\
\hline Mean Arrival Rate $(\lambda)$ (calls/unit time)& 1 to 80 \\
 \hline
 \end{tabular}
 \label{simulation}
 \end{table}
%\begin{figure*}[h!]
%  \begin{center}
%\scalebox{0.55}{\input{Txn_Sys.pstex_t}}
%\center \caption{Subcarrier allocation in Frequency and Time domain}
%  \label{Txn_sys}
%  \end{center}
%  \end{figure*}
% ======================================
\end{document}